\documentclass[aps, prl, reprint, superscriptaddress, amsmath, amssymb]{revtex4-2}

\usepackage{graphicx}% Include figure files
\usepackage{dcolumn}% Align table columns on decimal point
\usepackage{bm}% bold math
\date{\today}
\usepackage{xcolor}
\usepackage{calc}
\usepackage{subcaption}
\usepackage{caption}
\usepackage{kotex}
\usepackage{comment}
\usepackage{colortbl}
\allowdisplaybreaks 

\begin{document}

\setcounter{figure}{0} %Reset figure numbering
\global\long\def\thefigure{\arabic{figure}}%

\graphicspath{ {./figs_main/} }

\preprint{APS/123-QED}

\title{Unveiling three types of fermions in a nodal ring topological semimetal through magneto-optical transitions}

\author{Jiwon Jeon}
 \thanks{These authors contributed equally to this work.}
%\affiliation{%
% Natural Science Research Institute, University of Seoul, Seoul 02504, Korea
% }%
 \affiliation{%
 Physics Department, University of Seoul, Seoul 02504, Korea
 }%

\author{Taehyeok Kim}%
 \thanks{These authors contributed equally to this work.}
 \affiliation{%
 Department of Physics and Astronomy, Seoul National University, Seoul 08826, Korea
 }%

 \author{Jiho Jang}%
 \affiliation{%
 Department of Physics and Astronomy, Seoul National University, Seoul 08826, Korea
 }%

\author{Hoil Kim}%
 \affiliation{%
 Center for Artificial Low Dimensional Electronic Systems, Institute for Basic Science (IBS), Pohang 37673, Korea
 }%
  \affiliation{%
 Department of Physics, Pohang University of Science and Technology (POSTECH), Pohang 37673, Korea
 }%

\author{Mykhaylo Ozerov}%
 \affiliation{%
 National High Magnetic Field Laboratory, Tallahassee, Florida 32310, USA
 }%

\author{Jun Sung Kim}%
 \affiliation{%
 Center for Artificial Low Dimensional Electronic Systems, Institute for Basic Science (IBS), Pohang 37673, Korea
 }%
  \affiliation{%
 Department of Physics, Pohang University of Science and Technology (POSTECH), Pohang 37673, Korea
 }%

\author{Hongki Min}%
 \email{hmin@snu.ac.kr}
 \affiliation{%
 Department of Physics and Astronomy, Seoul National University, Seoul 08826, Korea
 }%

\author{Eunjip Choi}%
\email{echoi@uos.ac.kr}
 \affiliation{%
 Physics Department, University of Seoul, Seoul 02504, Korea
 }%

\date{\today}% It is always \today, today,

\begin{abstract}

We investigate the quasiparticles of a single nodal ring semimetal $\rm{SrAs}_3$ through axis-resolved magneto-optical measurements. We observe three types of Landau levels scaling as $\varepsilon \sim \sqrt{B}$, $\varepsilon \sim B^{2/3}$, and $\varepsilon \sim B$ that correspond to Dirac, semi-Dirac, and classical fermions, respectively. Through theoretical analysis, we identify the distinct origins of these three types of fermions present within the nodal ring. In particular, semi-Dirac fermions—a novel type of fermion that can give rise to a range of unique quantum phenomena—emerge from the endpoints of the nodal ring where the energy band disperses linearly along one direction and quadratically along the perpendicular direction, a feature not achievable in nodal point or line structures. The capacity of the nodal ring to simultaneously host multiple fermion types, including semi-Dirac fermions, establishes it as a valuable platform to expand the understanding of topological semimetals.

\end{abstract}

\maketitle

%\tableofcontents

Topological semimetals represent a class of materials characterized by conduction and valence bands that cross near the Fermi level. These band crossings, protected by symmetry, result in the formation of nodal points or nodal lines that host exotic quasiparticles, such as Dirac and Weyl fermions. These quasiparticles exhibit novel quantum phenomena, including the chiral anomaly, anomalous Hall effect, and axion electrodynamics \cite{Armitage2018, Lv2021, 10.1063/5.0038804}.

Nodal rings are a special form of band-crossing structure.
A nodal ring can exhibit, due to its closed and curved geometry, unique quantum phenomena such as Berry flux encirclement, weak antilocalization, and drumhead surface states \cite{Kim2022, Hosen2020, Rui2018, Zhao2023}.
In particular, a nodal ring can harbor a novel type of fermion, termed the semi-Dirac fermion \cite{Oroszlany2018}, which acts as Dirac fermions along one direction and classical fermions along the perpendicular direction. This hybrid property of semi-Dirac fermions leads to a range of novel quantum phenomena including exotic non-Fermi liquid states \cite{PhysRevLett.116.076803}, unconventional superconductivity \cite{PhysRevB.96.220503}, and unusual hydrodynamic transport \cite{PhysRevLett.120.196801}.
However, despite these significant theoretical insights, experimental investigations of nodal ring semimetals are limited compared to those of nodal point and nodal line semimetals.

\begin{figure}[htb!]
    \centering
    \includegraphics[width=\linewidth]{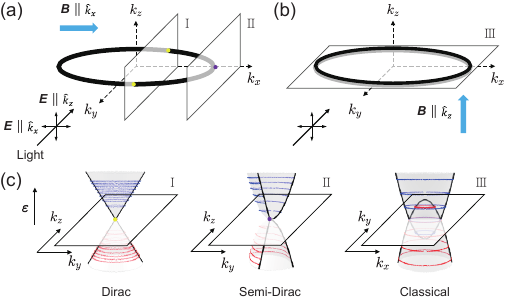}
    \caption{Schematic diagrams of magneto-optical measurements on a single nodal ring. A magnetic field is applied parallel to $k_x$ and $k_z$ in (a) and (b), respectively. Incident light is polarized along $k_x$ or $k_z$. (c) Energy dispersions within the three guiding planes I, II, and III shown in (a) and (b), which are linear (I), quadratic (III), and linear+quadratic (II). In (II), the energy bands are cut to illustrate the linear+quadratic dispersions clearly.
    The blue and red lines represent the LLs with distinct energy spacings, $\varepsilon_n\sim\sqrt{nB}$ for Dirac, $\varepsilon_n\sim\left[\left(n+{1\over 2}\right)B\right]^{2/3}$ for semi-Dirac, and $\varepsilon_n\sim \left(n+{1\over 2}\right)B$ for classical fermions, respectively, where $n$ is the LL index.
    }
    \label{Fig 1: Schematic picture of the ideal nodal ring}
\end{figure}

\begin{figure*}[htb!]
    \centering
    \includegraphics[width=\linewidth]{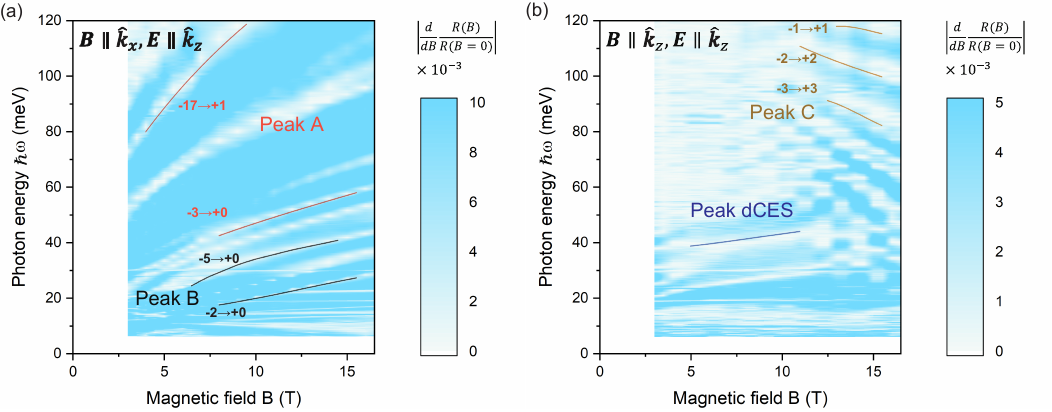}
    \caption{Experimental magneto-optical transition peaks measured in the (a) $(\bm{B}, \bm{E}) \parallel (\hat{\bm{k}}_x, \hat{\bm{k}}_z)$ mode and
    (b) $(\bm{B}, \bm{E}) \parallel (\hat{\bm{k}}_z, \hat{\bm{k}}_z)$ mode. 
    The derivative of reflectance ratio $\left|\frac{d}{dB} \frac{R(B)}{R(B=0)} \right|$ was employed in the 2D color plot to enhance visibility. 
    The shaded thick lines represent the regions exhibiting characteristic LL transition peaks corresponding to Peak A-C and Peak dCES. The LL indices involved in the optical transitions are labeled, as shown in Fig. S8(a) for Peak A, Fig. S8(b) for Peak B, and Fig. S4(a) for Peak C. For the high energy peak of Peak A, several LL transitions are involved, where $n=-17 \rightarrow +1$ is the representative one, as shown in Fig. S14.}
    \label{Fig 2: ...}
\end{figure*}

For probing fermions within topological semimetals, magneto-infrared spectroscopy has proven to be an effective approach. When subjected to a magnetic field, Dirac fermions governed by the linear energy dispersion exhibit Landau levels (LLs) that scale as $\varepsilon \sim \sqrt{B}$, in contrast to classical fermions with parabolic energy dispersion which display the standard $\varepsilon \sim B$ scaling. Notably, semi-Dirac fermions are predicted to manifest a new type of LL scaling, 
$\varepsilon \sim B^{2/3}$ due to the hybrid of linear and parabolic energy dispersions \cite{Banerjee2009, Sinha2022}. 
The LL scaling can be determined experimentally through the inter-LL optical transitions. 
While numerous LL transition measurements  have been conducted on topological semimetals \cite{PhysRevB.107.L241101, PhysRevMaterials.6.054204, doi:10.1126/sciadv.abj1076, PhysRevLett.124.176402, PhysRevB.104.L201115, PhysRevB.108.L241201, PhysRevB.108.L241201, Shao2020, Uykur2019, Shao2024}, those revealing the signature of semi-Dirac fermions remain sparse.
Therefore, the investigations of LL transitions and their corresponding LLs scaling behaviors are critical for unveiling the fermions emerging from a nodal ring.

In this work, we perform magneto-infrared measurements on $\rm{SrAs}_3$. $\rm{SrAs}_3$ is a recently identified nodal ring semimetal possessing two key features that make it particularly well-suited for optical studies \cite{Li2018, Hosen2020, Kim2022, Xu2017}. Firstly, $\rm{SrAs}_3$ features only a single nodal ring within the Brillouin zone (BZ), and secondly, its low-energy optical transitions arise solely from non-trivial bands, without interference from topologically trivial bands \cite{PhysRevB.95.045136, Hosen2020, Wang2024}.
The single nodal ring structure of $\rm{SrAs}_3$ enables controlled alignment of the magnetic field $\bm{B}$ and the electric field $\bm{E}$ of incident light along specific directions, such as parallel or perpendicular to the nodal ring axis.
This capability allows direction-resolved measurements of LL transitions, thus facilitating a comprehensive investigation of the fermions inherent in the nodal ring. 

We reveal that  LLs with the novel $\varepsilon \sim B^{2/3}$ scaling emerge in $\rm{SrAs}_3$, demonstrating the presence of semi-Dirac fermions in the nodal ring. Additionally, we observe two other LLs scaling as $\varepsilon \sim B^{1/2}$ and $\varepsilon \sim B$, arising from Dirac and classical fermions, respectively. 
We perform a theoretical analysis of the LLs and identify the origins of the three types of fermions.
The coexistence of these three types of fermions represents a unique property of the nodal ring, distinguishing it from nodal point or line structures.

\begin{figure*}
    \centering
    \includegraphics[width=\linewidth]{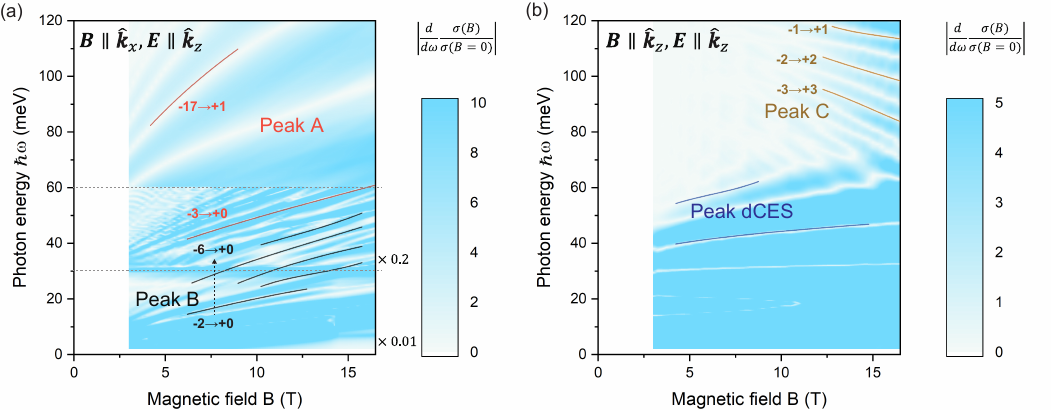}
    \caption{Theoretical magneto-optical transition peaks calculated in the (a) $(\bm{B}, \bm{E}) \parallel (\hat{\bm{k}}_x, \hat{\bm{k}}_z)$ mode and (b) $(\bm{B}, \bm{E}) \parallel (\hat{\bm{k}}_z, \hat{\bm{k}}_z)$ mode. The derivative $\left|\frac{d}{d\omega} \frac{\sigma(B)}{\sigma(B=0)} \right|$ was employed in the 2D color plot to enhance visibility. The shaded thick lines represent the regions exhibiting characteristic LL transition peaks corresponding to Peak A-C and Peak dCES, with the LL indices and colors in both modes matching those observed experimentally in Fig. \ref{Fig 2: ...}. For numerical calculations, we used the broadening term $\eta=4$ meV, except for $\hbar\omega<60$ meV in (a) where $\eta=1$ meV was used for visibility. The color intensity was adjusted  in the $30\ - \ 60$ meV range and $< 30$ meV range by multiplying the initial values by 0.2 and 0.01, respectively.}
    \label{Fig 3: ...}
\end{figure*}

Single crystals of $\rm{SrAs}_3$ were grown using the Bridgman method and their structural, transport, and electron band properties were extensively characterized \cite{Kim2022}. The optical reflectivity of $\rm{SrAs}_3$ was measured at zero magnetic field, and the corresponding optical transitions were theoretically analyzed in Ref. \cite{Jeon2023}.
The magneto-optical measurements were performed under Voigt geometry at $T=4.2 \ {\rm K}$.
Broadband reflectances $R(B)$ were taken from $B=0$ up to 17.45 T over the far- and mid-infrared ranges (7 meV$ - $400 meV) using a Fourier transform spectrometer and a bolometric detector.

Figure \ref{Fig 1: Schematic picture of the ideal nodal ring} 
illustrates the schematic configuration of magneto-optical measurements on a single nodal ring, including the directions of the magnetic field and light polarization. In Fig. \ref{Fig 1: Schematic picture of the ideal nodal ring}(a), a magnetic field is applied along $\bm{B} \parallel \hat{\bm{k}}_x$ which is parallel to the nodal ring plane. Incident light propagates along the $k_y$ direction with the electric field $\bm{E}$ polarized along $\bm{E} \parallel \hat{\bm{k}}_x$ or $\bm{E} \parallel \hat{\bm{k}}_z$. In Fig. \ref{Fig 1: Schematic picture of the ideal nodal ring}(b), the magnetic field is applied perpendicular to the nodal ring plane, $\bm{B} \parallel \hat{\bm{k}}_z$.  
These combinations of $\bm{B}$ and $\bm{E}$ yield four kinds of measurement modes, $(\bm{B}, \bm{E}) \parallel (\hat{\bm{k}}_z, \hat{\bm{k}}_z), \ (\hat{\bm{k}}_x, \hat{\bm{k}}_z), \ (\hat{\bm{k}}_z, \hat{\bm{k}}_x), \ {\rm and} \ (\hat{\bm{k}}_x, \hat{\bm{k}}_x)$.
Further details of the optical measurements are described in Fig. S1 of Supplemental Material (SM) Sec. I. 
In Figs. \ref{Fig 1: Schematic picture of the ideal nodal ring}(a) and \ref{Fig 1: Schematic picture of the ideal nodal ring}(b), we introduce three guiding planes I, II, and III that cross the nodal ring to visualize the energy band dispersions within them. These dispersions are linear (I), quadratic (III), and linear along one direction while quadratic along the perpendicular direction (II), as shown in Fig. 1(c). These three types of band structures are derived from a model Hamiltonian of the nodal ring, which will be described later in the theory section. 

Figure \ref{Fig 2: ...} shows the results of optical reflectance measurements for the two $(\bm{B}, \bm{E})$ modes with $\bm{E} \parallel \hat{\bm{k}}_z$, i.e, $(\bm{B}, \bm{E}) \parallel (\hat{\bm{k}}_x, \hat{\bm{k}}_z), \ {\rm and} \ (\hat{\bm{k}}_z, \hat{\bm{k}}_z)$
In the normalized reflectance $R(B)/R(B=0)$, multiple LL transitions emerge and evolve as $B$ is applied (see SM Sec. I). To enhance the visibility of the LL transitions,
we plot the derivative of the normalized reflectance in Fig. 2.
In the $(\bm{B}, \bm{E}) \parallel (\hat{\bm{k}}_x, \hat{\bm{k}}_z)$ mode, two classes of LL transitions, labeled as Peak A and Peak B, are observed. In Peak A (orange), LL transitions evolve within the $40 - 120$ meV ranges as $B$ is increased. In contrast, the LL transitions of Peak B (black) appear at significantly lower energy ranges, indicating distinct origins of Peak A and Peak B.
For the $(\bm{B}, \bm{E}) \parallel (\hat{\bm{k}}_z, \hat{\bm{k}}_z)$ mode, two classes of LL transitions, Peak C and Peak dCES, where dCES represents the disconnected constant energy surface as will be described later, are observed. The LL transitions of Peak C (golden) decrease as $B$ increases, contrary to Peak A and Peak B. The LL transitions of Peak dCES (blue) emerge at relatively lower energy ranges below 40 meV. 

To elucidate the origins of the four classes of LL transitions, Peak A-C and  Peak dCES, we perform a theoretical analysis of the magneto-optical transitions of $\rm{SrAs}_3$. 
$\rm{SrAs}_3$ is characterized by an elliptical nodal ring formed around the Y point in the BZ. Optical transitions near the Fermi energy are confined to this nodal ring, enabling the derivation of magneto-optical conductivity using an effective Hamiltonian \cite{Jeon2023} as 
\begin{equation} \label{eqn: SrAs3 Hamiltonian}
H = f_0 \sigma_0 s_0 + f_1 \sigma_1 s_0 + \hbar v_z k_z \sigma_2 s_0 + \Delta_\text{SOC} \sigma_3 s_3,
\end{equation}
where $f_0 = a_0 + a_{x} k_{x}^2 + a_{xy} k_{x} k_y + a_{y} k_{y}^2 + a_{z} k_{z}^2$  and $f_1 = b_0 + b_{x} k_{x}^2 + b_{xy} k_{x} k_{y} + b_{y} k_{y}^2 + b_{z} k_{z}^2$. The orbital and spin degrees of freedom are represented by Pauli matrices $\bm{\sigma}$ and $\bm{s}$, respectively. The nodal ring is located where $k_z=0$ and $f_1=0$, with its axis aligned along the $k_z$-direction. The first term describes the energy tilt that shifts the energy of the nodal ring by $f_0$, and the last term accounts for SOC opening an energy gap of $2 \Delta_\text{SOC}$, respectively. Further details of Eq. (1) are described in SM Sec. II.

To develop a qualitative understanding of the LL transitions, we first consider a simplified form of Eq. (\ref{eqn: SrAs3 Hamiltonian}), neglecting the effects of tilt, SOC, and ellipticity:
\begin{equation} \label{eqn: simplified Hamiltonian}
    H = \frac{\hbar^2}{2m}\left( k_x^2+k_y^2-k_0^2 \right) \sigma_1 + \hbar v_z k_z \sigma_2,
\end{equation}
where the scale of momentum and energy are set to the nodal ring radius $k_0$ and $\varepsilon_0 = \hbar^2 k_0^2 /2m$, respectively.

First, when the magnetic field is oriented perpendicular to the nodal ring axis, $\bm{B}\ ||\ \hat{\bm{k}}_x$ as shown in Fig. 1(a), LLs form in the 2D transverse momentum plane, with $k_x$ (parallel to the field) remaining a good quantum number. The magneto-optical conductivity sums contributions from all $k_x$ values, capturing the distinct  LLs of Dirac fermions and semi-Dirac fermions depicted in Fig. 1(c). At $k_x=0$, the Hamiltonian around zero energy is given by
\begin{equation}
    H \big|_{k_x = 0} \approx \hbar v_y \delta k_y \sigma_1 + \hbar v_z k_z \sigma_2,
\end{equation}
where $\delta k_y = k_y \mp k_{0}$ and $v_y=\hbar k_0 /m$, describing (anisotropic) Dirac fermions, whereas at $k_x= \pm k_0$,
\begin{equation}
    H \big| _{k_x = \pm k_0}  \approx \frac{\hbar^2 k_y^2}{2m} \sigma_1 + \hbar v_z k_z \sigma_2,
\end{equation}
describing semi-Dirac fermions.
As the energy moves away from zero, the constant energy surfaces of the two
Dirac bands located at $k_x = \pm k_0$ expand and eventually merge. At this point, the Dirac approximation is no longer applicable. As $k_x$ shifts from zero, the energy range where the Dirac approximation is valid ($\left|\hbar \omega\right| < \hbar^2 (k_0^2 - k_x^2) / 2m$) narrows, leading towards the semi-Dirac regime.
Note that the curved geometry of the nodal ring plays a crucial role in forming the semi-Dirac fermions: On the tangential planes at the endpoints of a nodal ring, the plane II in Fig. \ref{Fig 1: Schematic picture of the ideal nodal ring}, the effective Hamiltonian supports linear dispersion along one direction and quadratic dispersion along the other direction, 
as illustrated in Fig. \ref{Fig 1: Schematic picture of the ideal nodal ring}(c), 
which is not achieved in straight nodal lines.
When the tilt and SOC terms are introduced, the tilt term modifies LLs without altering their scaling behavior, while the SOC term induces a gap, placing Dirac LL transitions above the SOC-induced gap. Due to the tilt-induced energy shift, semi-Dirac LL transitions remain visible even below the SOC gap. The distinction allows for clear identification of the two LL transitions, Dirac and semi-Dirac, in their respective energy ranges. These effects of the tilt and SOC are confirmed by numerical calculations (see SM Sec. III) for the effective Hamiltonian of $\rm{SrAs}_3$, including a magnetic field into Eq. (1).

Second, when the magnetic field is applied parallel to the nodal ring axis as shown in Fig. 1(b), the simplified effective Hamiltonian in the $k_z=0$ plane is given by
\begin{equation}
    H \big|_{k_z = 0} \approx \frac{\hbar^2}{2m}\left( k_x^2 + k_y^2 -k_0^2 \right) \sigma_1,
\end{equation}
which resembles those of classical electron and hole gases. 
The LLs inherit the properties of the classical electron (hole) gas, starting from their origin $-\varepsilon_0 \  (\varepsilon_0)$ and increasing (decreasing) linearly with the magnetic  field. Thus, the LL transitions start from $2 \varepsilon_0$, the energy separation of the two gases at their origins, and decrease linearly as the LLs shift with $B$. As $k_z$ deviates from zero, the second term in Eq. (\ref{eqn: simplified Hamiltonian}) generates a gap of $2\hbar v_z k_z $ between the two gases, resulting in reduced  contributions in the low-energy range of interest. 
The tilt term modifies the effective masses of the electron and hole gases, respectively while preserving the linear scaling of the LLs. 
The SOC term introduces a gap between the two gases, while its effect on the LL transitions is negligible as they occur far above this gap. Notably, 
the electron and hole pockets split into smaller pockets since the tilt term deforms their constant energy surfaces (CESs) into distinct ellipses, creating crossing points. In addition, the SOC term disconnects the deformed CESs around the crossing points. The disconnected CESs lead to 
a class of LL transitions Peak dCES. See SM Sec. IV-1.

Figure \ref{Fig 3: ...} shows the magneto-optical conductivity obtained from the theoretical calculations for the effective Hamiltonian of $\rm{SrAs}_3$, including a magnetic field into Eq. (1). The LL transitions of Peak A and Peak B  are observed in the $(\bm{B}, \bm{E}) \parallel (\hat{\bm{k}}_x, \hat{\bm{k}}_z)$ mode, whereas those of Peak C and Peak dCES appear in the $(\bm{B}, \bm{E}) \parallel (\hat{\bm{k}}_z, \hat{\bm{k}}_z)$ mode, with the respective energy ranges also 
in agreement with the experimental results. Based on this agreement, we extract the LLs responsible for the four classes of LL transition peaks.

\begin{figure}[h]
    \centering
    \includegraphics[width=\linewidth]{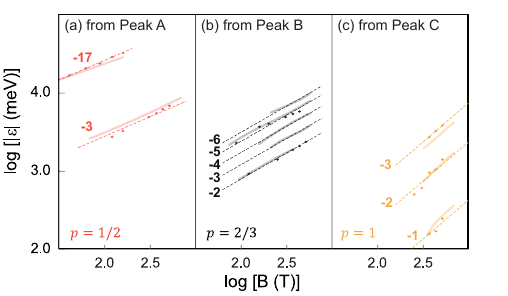}
    \caption{
   Logarithmic plots of LL energies as a function of magnetic field strength extracted from (a) Peak A, (b) Peak B, and (c) Peak C, with the respective LL indices indicated. Filled circles and solid lines represent experimental and theoretical results, respectively. Dashed lines guide the slopes of (a) 1/2, (b) 2/3, and (c) 1, corresponding to Dirac, semi-Dirac, and classical fermions, respectively.
    }
    \label{Fig 4: compare exp and cal fitting curve}
\end{figure}

Figure 4 presents the relationship between the extracted LL energies and magnetic field strength, using a logarithmic plot to highlight their scaling behaviors, in both the experimental and numerical data. 
In Fig. 4(a), with the magnetic field applied perpendicular to the nodal ring axis, $\bm{B} \parallel \hat{\bm{k}}_x$, the LLs energies exhibit the slope of 1/2 (extracted from Peak A) and 2/3 (Peak B), corresponding to Dirac and semi-Dirac fermions, respectively. In Fig. 4(b), with the field applied parallel to the nodal ring axis, $\bm{B} \parallel \hat{\bm{k}}_z$, the LLs exhibit the slope of 1 (Peak C) aligning with classical fermions.
The experimental and numerical results consistently demonstrate that the LLs represent the three types of fermions, Dirac, semi-Dirac, and classical fermions.
The LLs extracted from Peak dCES deviate from the simple scaling behaviors due to complex band dispersions of the disconnected electron and hole pocket. The derivation of LLs from the LL transition peaks is described in SM Sec. IV.  

To this point, we focused on the results of the two $(\bm{B}, \bm{E})$ modes with $\bm{E} \parallel \hat{\bm{k}}_z$ polarization. For the other polarization $\bm{E} \parallel \hat{\bm{k}}_x$, the LLs exhibit the same scaling behaviors as in $\bm{E} \parallel \hat{\bm{k}}_z$, while the corresponding LL transitions are notably weaker in the peaks' intensities, especially for Peak B and Peak C, since the corresponding effective velocity $v_x$ diminishes. See SM Sec. V. These findings demonstrate that an axis-resolved study is essential for a comprehensive understanding of the LL transitions of the nodal ring.

To conclude, we identified the complete, direction-resolved LL transitions of $\rm{SrAs}_3$ through magneto-optical measurements. The four kinds of $(\bm{B}, \bm{E})$ measurement modes, enabled by the unique single nodal ring structure, were critical in revealing them. The LL transitions elucidated two key properties of the nodal ring: First, the nodal ring hosts three distinct types of fermions simultaneously $\text{--}$ Dirac, semi-Dirac, and classical fermions $\text{--}$ distinguishing it from nodal point or line structures which typically exhibit only one or two fermion types. See SM Sec. VI. Their coexistence results from the unique geometry of the nodal ring, which, as shown in Fig. \ref{Fig 1: Schematic picture of the ideal nodal ring}, supports linear, linear+quadratic, and quadratic energy dispersions within a single structure. The three types of fermions are well-separated from each other in their excitation energies and/or in the $(\bm{B}, \bm{E})$ configurations as demonstrated in Fig. \ref{Fig 4: compare exp and cal fitting curve}, providing an opportunity to experimentally probe and compare their properties within the same material.
Second, we emphasize the emergence of semi-Dirac fermions in the single nodal ring. In $\rm{SrAs}_3$, the curved geometry of the nodal ring ensures the existence of endpoints and, thereby, semi-Dirac fermions arising from them. The semi-Dirac fermions can be accessed for any orientation of $\bm{B}$, provided that $\bm{B}$ lies within the nodal ring plane, facilitating experimental feasibility that is not achieved in other nodal line semimetals possessing straight lines. The semi-Dirac fermions bridge the gap between Dirac and classical fermions, serving as an intermediary state as demonstrated by their LL scaling power (2/3) lying between 1/2 (Dirac) and 1 (classical). The ease of access to semi-Dirac fermions provides a framework for experimentally investigating diverse quantum properties beyond the unique LL scaling behavior.
Overall, our findings in $\rm{SrAs}_3$ establish that a nodal ring serves as a valuable platform for advancing the understanding of topological semimetals.

$Note\ added$: Recently, a $\hbar\omega \sim B^{2/3}$ scaling was reported in ZrSiS \cite{Shao2024}, arising from certain crossing points of multiple nodal lines, in contrast to the single nodal ring of the current work.

\begin{acknowledgments}
The work at UOS was supported by the NRF grants funded by the Korea government (Grant No. 2021R1A2C1009073). The work at SNU was supported by the NRF grants funded by the Korea government (MSIT) (Grant No. 2023R1A2C1005996),  the Creative-Pioneering Researchers Program through SNU, and the Center for Theoretical Physics. The work at POSTECH was supported by the Institute for Basic Science (no. IBS-R014-D1), by the NRF of Korea government through the Basic Science Research Program (Grant No. NRF-2022R1A2C3009731), and the Max Planck POSTECH/Korea Research Initiative (Grant No. 2022M3H4A1A04074153).
The NHMFL is supported by NSF through NSF/DMR-2128556 and the State of Florida.
\end{acknowledgments}

\nocite{*}

\bibliographystyle{apsrev4-1}
\bibliography{main}% Produces the bibliography via BibTeX.

\end{document}

\title{Supplemental Material: Unveiling three types of fermions\\
in a nodal ring topological semimetal through magneto-optical transitions}

\author{Jiwon Jeon}
 \thanks{These authors contributed equally to this work.}
%\affiliation{%
% Natural Science Research Institute, University of Seoul, Seoul 02504, Korea
% }%
 \affiliation{%
 Physics Department, University of Seoul, Seoul 02504, Korea
 }%

\author{Taehyeok Kim}%
 \thanks{These authors contributed equally to this work.}
 \affiliation{%
 Department of Physics and Astronomy, Seoul National University, Seoul 08826, Korea
 }%

 \author{Jiho Jang}%
 \affiliation{%
 Department of Physics and Astronomy, Seoul National University, Seoul 08826, Korea
 }%

\author{Hoil Kim}%
 \affiliation{%
 Center for Artificial Low Dimensional Electronic Systems, Institute for Basic Science (IBS), Pohang 37673, Korea
 }%
  \affiliation{%
 Department of Physics, Pohang University of Science and Technology (POSTECH), Pohang 37673, Korea
 }%

\author{Mykhaylo Ozerov}%
 \affiliation{%
 National High Magnetic Field Laboratory, Tallahassee, Florida 32310, USA
 }%

\author{Jun Sung Kim}%
 \affiliation{%
 Center for Artificial Low Dimensional Electronic Systems, Institute for Basic Science (IBS), Pohang 37673, Korea
 }%
  \affiliation{%
 Department of Physics, Pohang University of Science and Technology (POSTECH), Pohang 37673, Korea
 }%

\author{Hongki Min}%
 \email{hmin@snu.ac.kr}
 \affiliation{%
 Department of Physics and Astronomy, Seoul National University, Seoul 08826, Korea
 }%

\author{Eunjip Choi}%
\email{echoi@uos.ac.kr}
 \affiliation{%
 Physics Department, University of Seoul, Seoul 02504, Korea
 }%

\date{\today}% It is always \today, today,
\newcommand{\beginsupplement}{%
        \setcounter{table}{0}
        \renewcommand{\thetable}{S\arabic{table}}%
        \setcounter{figure}{0}
        \renewcommand{\thefigure}{S\arabic{figure}}%
     }
\captionsetup{justification=raggedright, size=small} 

%---------------------------------------
%\end{comment}


\maketitle
\beginsupplement
\graphicspath{ {./figs_main/} }
\tableofcontents

\newpage

\section{\uppercase\expandafter{\romannumeral1}. Experimental details}
\label{sec:SM-Experimental_details}

\begin{figure}[ht]
    \centering
    \includegraphics[width=\linewidth]{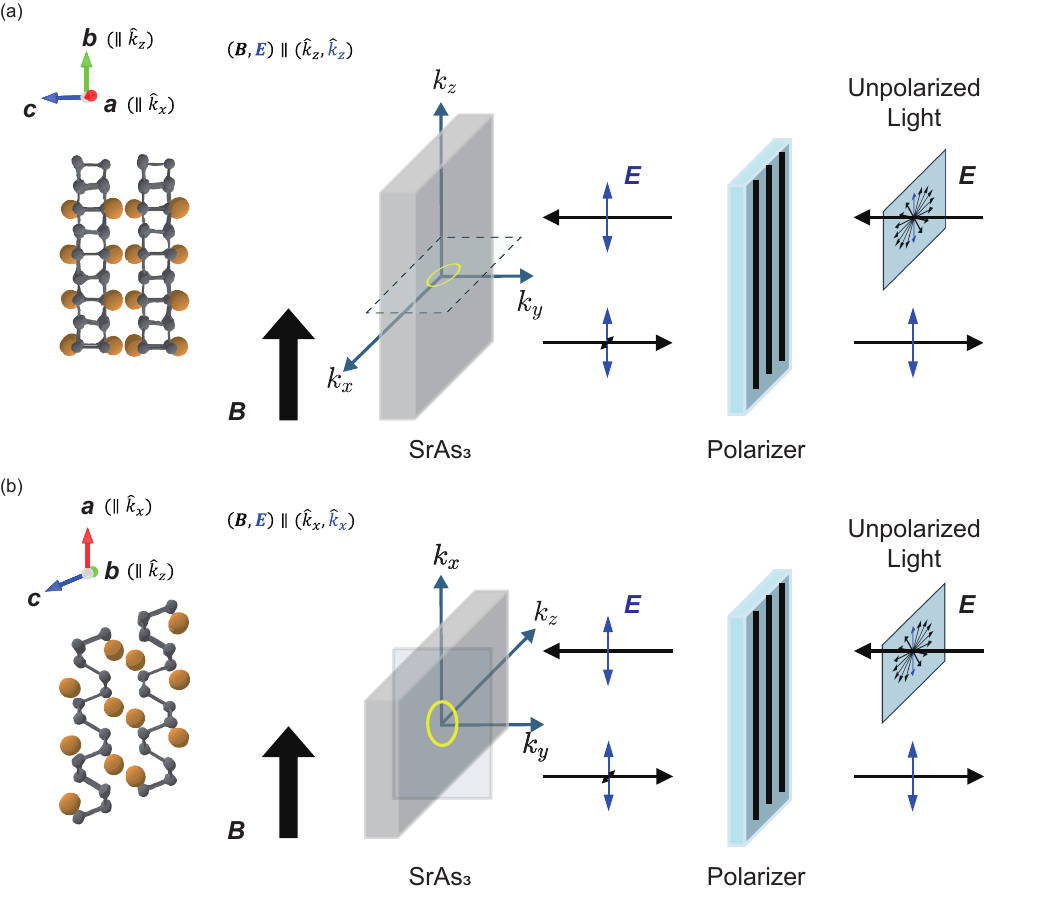}
    \caption{Configurations of the magneto-reflection measurements}
    \label{SFig 1: Schematic picture of the experiment}
\end{figure}

Figure \ref{SFig 1: Schematic picture of the experiment} shows the magneto-infrared measurement configurations of SrAs$_3$.
In Fig. \ref{SFig 1: Schematic picture of the experiment}(a), $\bm{B}$ is applied along the $k_z$-axis and $\bm{E}$ is polarized along the $k_z$-axis,  ($\bm{B}$, $\bm{E}$) $\parallel$ ($\hat{\bm{k}}_z$, $\hat{\bm{k}}_z$). The nodal ring lies within the $k_x$-$k_y$ plane. 
In Fig. \ref{SFig 1: Schematic picture of the experiment}(b), the sample is rotated with respect to Fig. \ref{SFig 1: Schematic picture of the experiment}(a) so that $\bm{B}$ is applied along the $k_x$-axis. The incident light is polarized along the $k_x$-axis. The nodal ring lies within the $k_x$-$k_z$ plane. 
In both Figs. \ref{SFig 1: Schematic picture of the experiment}(a) and \ref{SFig 1: Schematic picture of the experiment}(b), the reflected light is passed through the polarizer once more and subsequently measured by an optical detector. 

\clearpage

\begin{figure}[ht]
    \centering
    \includegraphics[width = \linewidth]{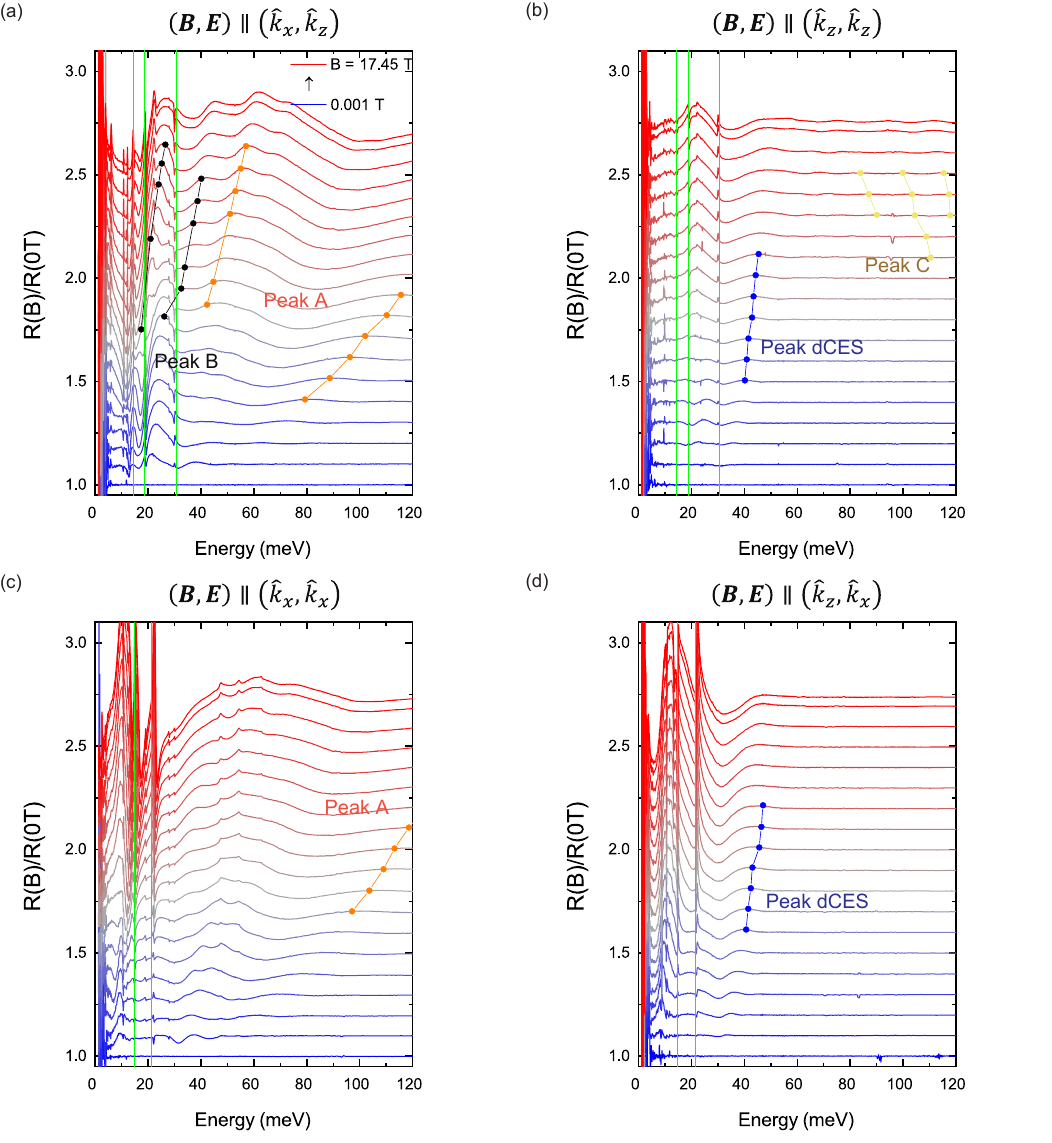}
    \caption{R(B)/R(0) data measured for the four ($\bm{B}$, $\bm{E}$) modes.}
    \label{SFig 2: R(H)/R(0T) data each measurement modes}
\end{figure}

Figure \ref{SFig 2: R(H)/R(0T) data each measurement modes} presents the reflectivity ratio $R(B)/R(0)$ for the four ($\bm{B}$, $\bm{E}$) measurement modes. Here, the maxima in $R(B)/R(0)$ are identified as LL transition energies.
Peak A (orange), Peak B (black), Peak C (golden), and Peak dCES (blue) show the LL transitions, while the green lines represent optical phonon peaks.

\clearpage

\section{\uppercase\expandafter{\romannumeral2}. The effective model of $\rm{SrAs}_3$}

A nodal ring presents a rich spectrum of Landau-level structures, which are crucial for understanding its electronic properties. When a magnetic field is applied, the two-dimensional (2D) momentum plane perpendicular to the field becomes quantized, leading to the formation of Landau levels. Within this 2D plane, the projection of the nodal ring reveals distinct features, such as Dirac fermions, semi-Dirac fermions, and classical fermions. These features demonstrate characteristic dependencies between the magnetic field strength and the Landau level energy.

Since the low-energy optical transitions are confined to the nodal ring structure located at the Y point in the Brillouin zone (BZ), we use an effective Hamiltonian of the nodal ring given by
\begin{equation} \label{eqn: Hamiltonian with tilt and SOC}
    H = f_0 \sigma_0 s_0 + f_1 \sigma_1 s_0 + \hbar v_z k_z \sigma_2 s_0 + \Delta_\text{SOC} \sigma_3 s_3,
\end{equation}
where $f_0 = a_0 + a_x k_x^2 + a_y k_y^2 + a_{xy} k_x k_y + a_z k_z^2$, $f_1 = b_0 + b_x k_x^2 + b_y k_y^2 + b_{xy} k_x k_y + b_z k_z^2$, and $\bm{\sigma}$ and $\bm{s}$ denote the orbital and spin degrees of freedom, respectively. The first term describes the energy tilt term, shifting the energy by $f_0$, and the last term accounts for SOC, opening an energy gap $2 \Delta_\text{SOC}$ along the nodal ring. The model parameters are summarized in Tab.~\ref{tab:parameters} \cite{Jeon2023}.

\begin{table}[h]
\renewcommand{\arraystretch}{1.0}
\begin{tabular*}{0.9\linewidth}{@{\extracolsep{\fill}} cccccc }
\hline \hline
$a_{0}$ ($\mathrm{eV}$) & $a_{x}$ ($\mathrm{eV \cdot \AA^{2}}$) & $a_{xy}$ ($\mathrm{eV \cdot \AA^{2}}$) & $a_{y}$ ($\mathrm{eV \cdot \AA^{2}}$) & $a_{z}$ ($\mathrm{eV \cdot \AA^{2}}$) & $\Delta_{\mathrm{SOC}}$ (eV) \\
  -0.0448 & 2.744 & -0.319 & 12.736  & -7.230 & 0.015 \\ \hline 
$b_{0}$ ($\mathrm{eV}$) & $b_{x}$ ($\mathrm{eV \cdot \AA^{2}}$) & $b_{xy}$ ($\mathrm{eV \cdot \AA^{2}}$) & $b_{y}$ ($\mathrm{eV \cdot \AA^{2}}$) & $b_{z}$ ($\mathrm{eV \cdot \AA^{2}}$) & $\hbar v_z$ ($\mathrm{eV \cdot \AA^{1}}$) \\
  0.0645 & -12.849 & -3.091 & -15.741 & 9.215 & 2.292 \\ \hline \hline
\end{tabular*}
\caption{The parameters for the Hamiltonian in Eq. (\ref{eqn: Hamiltonian with tilt and SOC}).}
\label{tab:parameters}
\end{table}

To understand the electronic structure of a nodal ring, we first consider a simplified nodal ring model, neglecting the effects of tilt, SOC, and ellipticity. The model Hamiltonian describing the simplified nodal ring is given by
\begin{equation} \label{eqn: model Hamiltonian of the simplified nodal ring}
H=\frac{\hbar^2}{2m} \left(k_x^2 + k_y^2 -k_0^2\right)\sigma_1 + \hbar v_z k_z \sigma_2.
\end{equation}
%This Hamiltonian describes a nodal ring lying at the $k_x$-$k_y$ plane with the radius $k_0$. 
%The nodal ring forms where the energy $E = 0$. 
The eigenvalues are expressed as 
\begin{equation} \label{eqn: Energy} 
\varepsilon = \pm  \sqrt{\left[ \frac{\hbar}{2m} \left( k_x^2 + k_y^2 - k_0^2\right) \right]^2 + \left(\hbar v_z k_z \right)^2}.
\end{equation}
A nodal line structure arises when the conditions $k_x^2 + k_y^2 = k_0^2$ and $k_z = 0$ are satisfied, forming a circular nodal ring in the $k_x$-$k_y$ plane with the radius of $k_0$ at energy $\varepsilon=0$. \\

The presence of a magnetic field results in the quantization of transverse momentum space, which leads to the formation of Landau levels. We will consider two major magnetic field directions: one with the magnetic field aligned parallel to the nodal ring axis ($\bm{B}\parallel\hat{\bm{k}}_z$), and the other with the magnetic field aligned perpendicular to the nodal ring axis ($\bm{B}\parallel\hat{\bm{k}}_x$). In the nodal ring, three distinct types of in-plane effective Hamiltonians emerge: classical fermions, Dirac fermions, and semi-Dirac fermions. When the magnetic field is oriented along the $k_z$-axis, the effective Hamiltonian at $k_z=0$ corresponds to classical electron and hole gases, originating from $\varepsilon=-\varepsilon_0$ and $+\varepsilon_0$, respectively, where $\varepsilon_0=\hbar^2 k_0^2 / 2m$:
\begin{equation} \label{eqn: effective Hamiltonian B//z}
    H(k_x, k_y, k_z=0) = \frac{\hbar^2}{2m} \left(k_x^2 + k_y^2 -k_0^2\right)\sigma_1.
\end{equation}
On the other hand, when the magnetic field is aligned parallel to the $k_x$-axis, the effective Hamiltonian when $\left| k_x \right| < k_0$ resembles that of Dirac fermions:
\begin{equation}\label{eqn: effective Hamiltonian B//x-Dirac}
    \begin{aligned}
        H(|k_x|<k_0, k_y, k_z) =&\, \frac{\hbar^2}{2m} \left( k_y^2-k_{y0}^2 \right) \sigma_1 + \hbar              v_z k_z \sigma_2  \\
        \approx& \frac{\hbar^2 k_{y0}}{m}\delta k_y \sigma_1 + \hbar v_z k_z \sigma_2,
    \end{aligned}
\end{equation}
where $\delta k_y=k_y-k_{y0}$ and $k_{y0}=\pm \sqrt{k_0^2-k_x^2}$. The nodal ring crosses at $k_y=k_{y0}$, and the second line in Eq. (\ref{eqn: effective Hamiltonian B//x-Dirac})  describes the effective Hamiltonian around the nodal points. When $k_x=\pm k_0$, the effective Hamiltonian resembles that of semi-Dirac fermions:
\begin{equation}\label{eqn: effective Hamiltonian B//x-Semi-Dirac}
 H(k_x=\pm k_0, k_y, k_z) =\frac{\hbar^2k_y^2}{2m} \sigma_1 + \hbar v_z k_z \sigma_2.
\end{equation}
These types of fermions are distinguished by their unique scaling behavior of the Landau levels on the magnetic field, as will be explained in the next section. Even though the tilt and SOC terms are included, the power law of Landau-level energies for the magnetic field remains valid when appropriate corrections are applied.

\clearpage

\section{\uppercase\expandafter{\romannumeral3}. Analysis of the Landau levels}

When a magnetic field is applied, Landau levels emerge in the momentum plane perpendicular to the field. These levels can be derived from the Hamiltonian using the ladder operators defined by
\begin{equation}\label{eqn: ladder operator}
    a = \frac{l_B}{\sqrt{2}\hbar}\left( \Pi_x + i \Pi_y\right),\ a^\dagger = \frac{l_B}{\sqrt{2}\hbar}\left( \Pi_x - i \Pi_y\right),
\end{equation}
where $l_B=\sqrt{\hbar c/eB}$ represents the magnetic length. The momentum operators in the Hamiltonian are substituted by the ladder operators using Eq. (\ref{eqn: ladder operator}). We construct the Hamiltonian matrix for each orbital and spin in terms of the number states $\left| n \right>$, which are eigenstates of the number operator $n=a^\dagger a$. In model calculations, it is impractical to consider an infinite number of Landau levels, so we handle a finite number of states, ensuring they are sufficient to describe the region where the optical transitions occur. \\

The Onsager quantization rule also gives us insight into the physical interpretation of the Landau levels in the semiclassical approach. The Onsager quantization rule states that the area enclosed by the constant energy surface is quantized according to the following relation:
 \begin{equation} \label{eqn: Onsager quantization rule}
 S=\frac{2\pi} {l_B^2} \left(n+\frac{1}{2} - \gamma \right) \quad \left(n=0, 1, 2, ...\right),
 \end{equation}
where 1/2 is a Maslov index and $\gamma$ is the Berry phase divided by $2\pi$: 0 in conventional materials, and 1/2 in topological materials with $\pi$-Berry phase such as graphene. The Onsager quantization rule is a powerful tool for making a physical interpretation of Landau levels, even when an analytic solution of Landau levels is unavailable.

The relationship between the Landau levels and the magnetic field follows a scaling determined by the exponents of the in-plane dispersions perpendicular to the magnetic field. Along the $i$-direction, assume that $\varepsilon \sim k_i^p$ with the exponent $p$, implying that the $k_i$ momentum changes proportionally to the $1/p$-power as the energy varies. Since the area $S \sim \varepsilon^{1/p+1/q}$, using the Onsager quantization rule, the magnetic field dependence of the Landau-level energy follows a scaling law as
\begin{equation}
    \varepsilon_n \sim \left[ \left( n+ \frac{1}{2}- \gamma \right) B  \right]^{pq/(p+q)},
\end{equation}
where $p$ and $q$ are the exponents of the dispersion along each direction when the magnetic field is absent. The exponents are given by $(p, q)=(2, 2)$, $(1, 1)$, and $(2, 1)$ in classical fermions with parabolic dispersion, Dirac fermions with linear dispersion, and semi-Dirac fermions with linear dispersion along one direction and parabolic dispersion along the other direction, respectively. The Berry phase correction $\gamma$ is 0 for classical fermions and semi-Dirac fermions, whereas 1/2 for Dirac fermions. In summary, the Landau-level energies obey the following scaling behaviors for each system:
\begin{equation}
    \begin{aligned}
        \varepsilon_{n, \text{classical}} &\sim \left(n+\frac{1}{2} \right) B, \\
        \varepsilon_{n, \text{Dirac}} & \sim \sqrt{nB}, \\
        \varepsilon_{n, \text{semi-Dirac}} & \sim \left[ \left(n+\frac{1}{2} \right) B \right]^{2/3}.
    \end{aligned}
\end{equation}
These scaling behaviors can be observed in the magneto-optical transition energies, reflecting the underlying electronic structures of the nodal ring. The Landau levels of the nodal ring have been studied when the magnetic field is aligned parallel and perpendicular to the nodal ring axis \cite{Montambaux2009, Laszlo2018, Duan2020, Sun2020, Asafov2024}.

\subsection{\uppercase\expandafter{\romannumeral3}-1. Magnetic field parallel to the nodal ring axis ($\bm{B}\parallel\hat{\bm{k}}_z$)}

\textbf{Simplified nodal ring model}: Figure \ref{figs: Landau levels, B//z} depicts the evolution of Landau levels for the simplified model under a magnetic field aligned with the nodal ring axis, $\bm{B} \parallel \hat{\bm{k}}_z$. At $k_z=0$, the levels divide into two distinct bands: electron gas and hole gas. Each band originates at $\varepsilon = -\varepsilon_0$ or $+\varepsilon_0$ and exhibits a linear dependence on the magnetic field, characteristic of electron or hole gases [Fig. \ref{figs: Landau levels, B//z}(a)]. When $k_z \neq 0$, a gap emerges between the electron and hole bands [Fig. \ref{figs: Landau levels, B//z}(b)], which widens as $k_z$ increases [Fig. \ref{figs: Landau levels, B//z}(c)]. This widening gap further separates the bands, shifting them beyond the region of interest and diminishing their contribution to magneto-optical conductivity.

\begin{figure}[htb!]
    \centering
    \includegraphics[width=\linewidth]{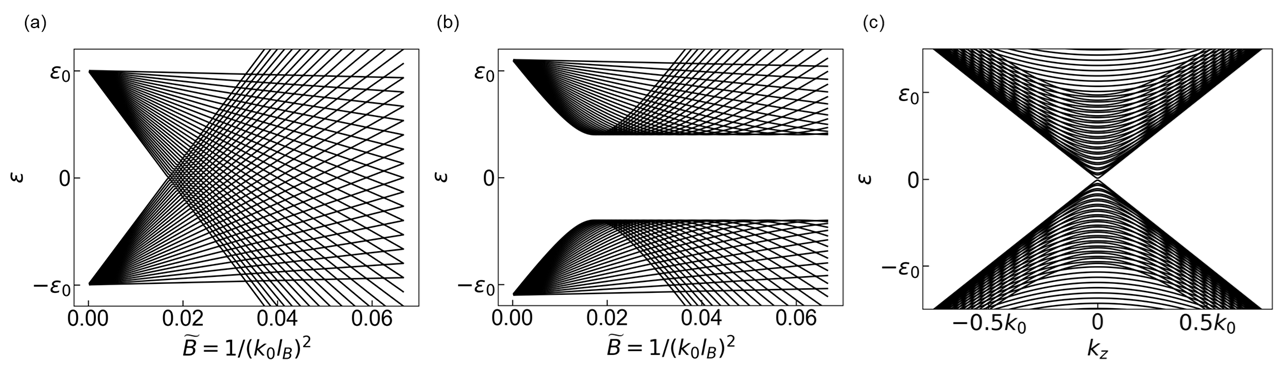}
    \caption{Landau levels of the simplified nodal ring model when $\bm{B}\parallel\hat{\bm{k}}_z$. Here, $v_z = \hbar k_0 / m$ is used and Landau levels are obtained numerically. For visibility, only 60 Landau levels are depicted. (a) At $k_z=0$, the Landau levels are linear for the magnitude of the magnetic field. (b) For $k_z = 0.2 k_0$, a gap appears between the Landau levels corresponding to the electron and hole gases. (c) Landau levels as a function of $k_z$. The magnetic field is set to $\tilde{B}=1/(k_0l_B)^2=0.03$. As $k_z$ deviates from zero, the gap between electron gas and hole gas becomes larger. }
    \label{figs: Landau levels, B//z}
\end{figure}

{\bf SrAs$_3$ model}: Figure \ref{figs: Landau levels, tilt & SOC, B//z} shows the Landau levels for the SrAs$_3$ model described in Eq. (\ref{eqn: Hamiltonian with tilt and SOC}) when $\bm{B} \parallel \hat{\bm{k}}_z$, including the effects of tilt and SOC. These levels are divided into two distinct regions, corresponding to classical fermions and disconnected constant energy surfaces. Beyond the energy range of $-40$ meV to $15$ meV, the Landau levels maintain a linear relationship with the magnetic field. With the inclusion of tilt, the energy of the electron and hole gases shifts by $f_0$. While the slopes of the Landau levels, determined by the inverse of the effective mass, are modified, their linear dependence on the magnetic field remains a defining feature originating from the electron and hole gases.

\begin{figure}[htb!]
    \centering
    \includegraphics[width=\linewidth]{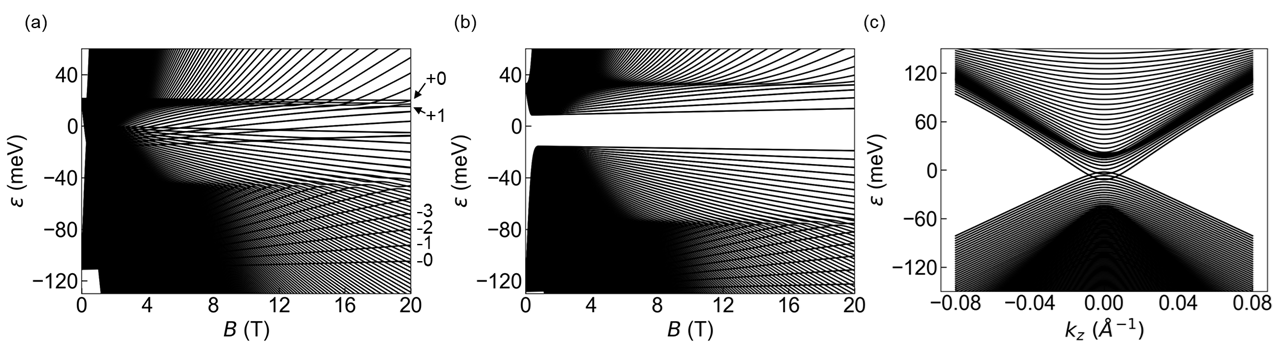}
    \caption{Landau levels of the SrAs$_3$ model when $\bm{B} \parallel \hat{\bm{k}}_z$: (a) $k_z=0$ and (b) $k_z = 0.02 \, \rm \AA^{-1}$. (c) Landau levels as a function of $k_z$ when $B=10$ T. In (a), LL indices corresponding to classical fermions are labeled.}
    \label{figs: Landau levels, tilt & SOC, B//z}
\end{figure}

\begin{figure}[htb!]
    \centering
    \includegraphics[width=\linewidth]{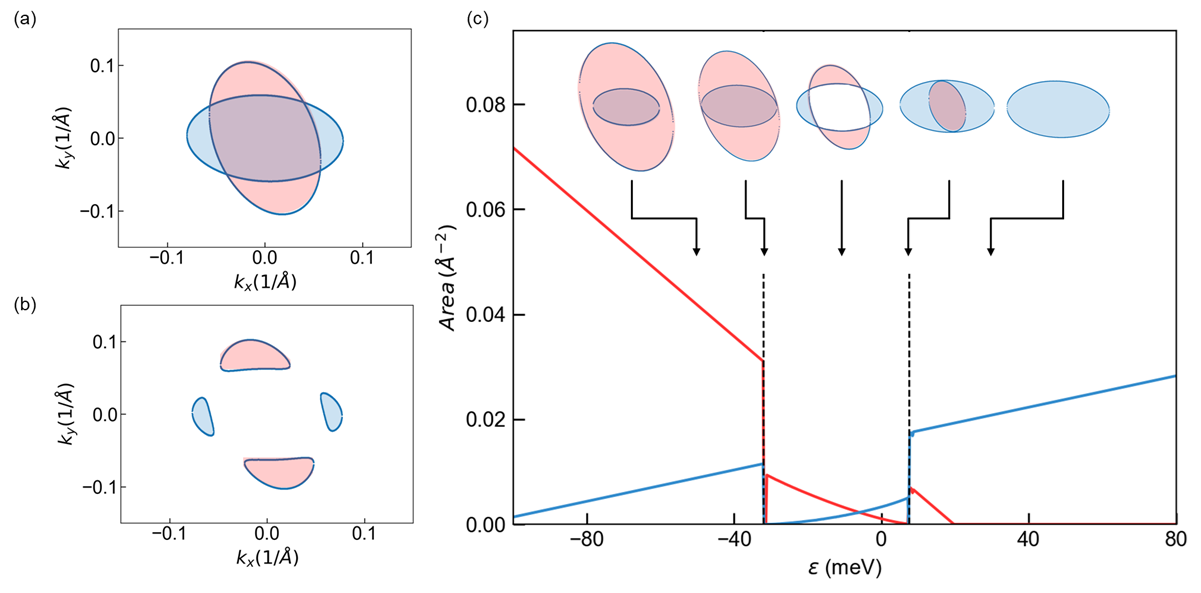}
    \caption{The constant energy surface at $-10$ meV when $k_z=0$ (a) without the SOC term and (b) with the SOC term. The tilt and SOC terms lead to disconnected constant energy surfaces, forming new Landau levels. The electron and hole pockets are depicted in blue and red, respectively. (c) The area enclosed by the constant energy surfaces, with dashed lines marking the two energy levels where the constant energy surfaces become disconnected. Here, we used the SrAs$_3$ model neglecting the SOC term to obtain the constant energy surface. Over the energy range of $-32$ to $8$ meV, we plot the area of a single electron or hole pocket. The schematic diagrams illustrating the evolution of the constant energy surface without the SOC term are shown above.}
    \label{fig: Fermi surface disconnection}
\end{figure}

Within the disconnected constant energy surface region ($-47$ meV to $22$ meV), the Landau levels evolve as the tilt and SOC terms disconnect the constant energy surfaces. Figure \ref{fig: Fermi surface disconnection} illustrates the constant energy surfaces and their corresponding enclosed areas with and without the SOC term. The tilt term deforms the electron and hole gases into distinct elliptical shapes, creating crossing points between the two ellipses. The SOC term further disconnects the constant energy surfaces, leading to the formation of smaller electron and hole pockets. The Onsager quantization rule, applied to the areas shown in Fig. \ref{fig: Fermi surface disconnection}(c), reveals that when the constant energy surfaces are disconnected, the areas change abruptly, resulting in new Landau levels originating from the smaller pockets. These Landau levels rise and fall for the electron and hole pockets, respectively.

\subsection{\uppercase\expandafter{\romannumeral3}-2. Magnetic field perpendicular to the nodal ring axis ($\bm{B}\parallel\hat{\bm{k}}_x$)}

\begin{figure}[htb!]
    \centering
    \includegraphics[width=\linewidth]{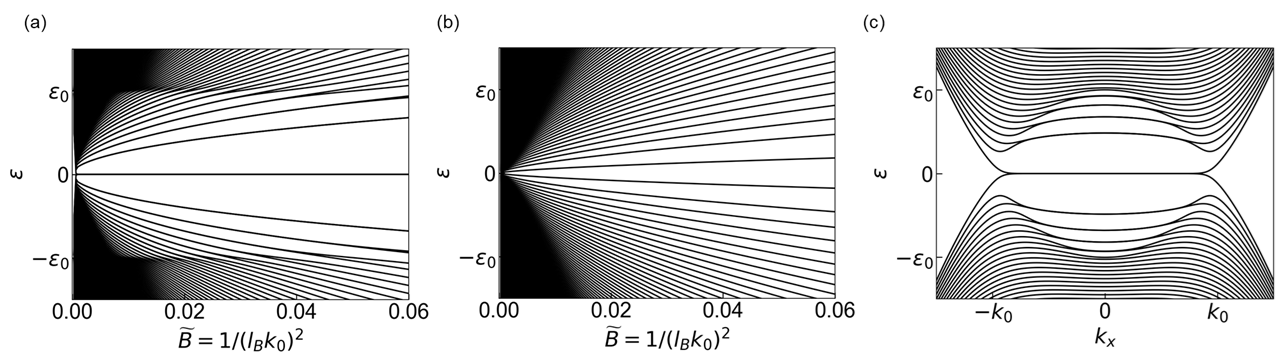}

\caption{Landau levels of the simplified nodal ring model when $\bm{B} \parallel \hat{\bm{k}}_x$. Here, $v_z = \hbar k_0 / m$ is used and Landau levels are obtained numerically. (a) When $k_x = 0$, the Landau levels follow $\varepsilon \sim \pm \sqrt{nB}$. In the region $|\varepsilon| > \varepsilon_0$, the Dirac-like effective Hamiltonian no longer applies, because the constant energy surfaces of the two Dirac points merge in this energy range. (b) When $k_x = k_0$, the Landau levels follow $\varepsilon \sim \pm \left[\left(n + 1/2\right)B\right]^{2/3}$, characteristic of semi-Dirac fermions. (c) Landau levels as a function of $k_x$, where the magnetic field strength is set to $\tilde{B}=1/(k_0l_B)^2=0.03$.}
\label{figs: Landau levels, B//x}
\end{figure}

{\bf Simplified nodal ring model}: When a magnetic field is applied along the $k_x$-axis, the Landau levels of the simplified nodal ring model exhibit a $k_x$-dependent evolution, reflecting the properties of the underlying Hamiltonian in zero field. As shown in Fig.~\ref{figs: Landau levels, B//x}(a), Dirac-like behavior dominates for $\varepsilon_n \sim B^{1/2}$, with zero-energy modes originating from the $\pi$-Berry phase. Figure \ref{figs: Landau levels, B//x}(b) reveals semi-Dirac behavior, characterized by $\varepsilon \sim B^{2/3}$. In Fig.~\ref{figs: Landau levels, B//x}(c), both Dirac and semi-Dirac characteristics coexist as a function of $k_x$, with their boundary defined by $\left| \varepsilon \right| = \frac{\hbar^2}{2m} \left( k_0^2-k_x^2 \right)$. Summing contributions across $k_x$, the magneto-optical conductivity exhibits a mixed signature of Dirac and semi-Dirac fermions, depending on the transition energy.

 \begin{figure}[htb!]
     \centering
     \includegraphics[width=\linewidth]{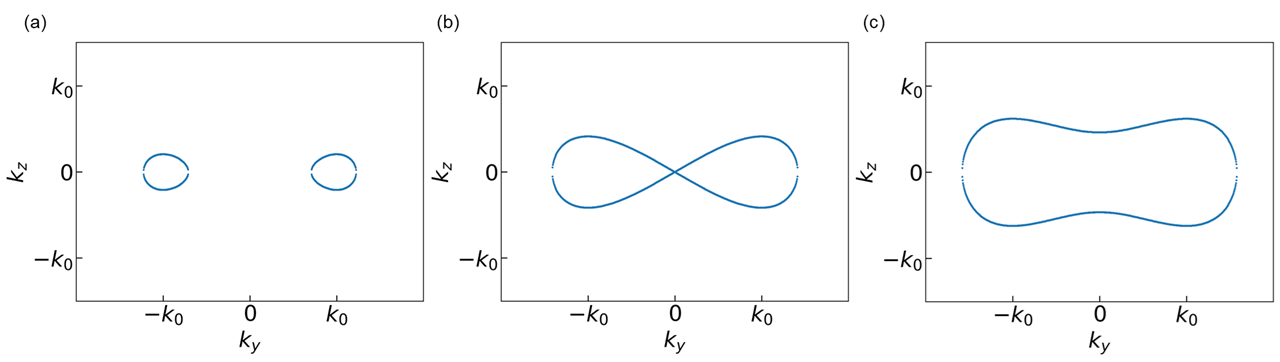}
     \caption{The constant energy surfaces of the simplified nodal ring model at $k_x=0$ for different energies: (a) $0.5\varepsilon_0$ (b) $\varepsilon_0$, and (c) $1.5\varepsilon_0$.}
     \label{fig: Fermi surface merge}
 \end{figure}

Figure \ref{fig: Fermi surface merge} depicts the constant energy surfaces for $\varepsilon=0.5 \varepsilon_0$, $\varepsilon_0$, and $1.5\varepsilon_0$. Near $k_x=0$ and $\varepsilon=0$, the system behaves like Dirac fermions. As the energy moves away from zero, the constant energy surfaces expand and merge at $\varepsilon = \pm \varepsilon_0$, marking the breakdown of the Dirac approximation. The constant energy surfaces coalesce at $\left| \varepsilon \right| = \frac{\hbar^2}{2m} \left( k_0^2-k_x^2 \right)$ as $k_x$ moves away from 0, marking the boundary between the Dirac-like and semi-Dirac-like regions. When $k_x = \pm k_0$, the system exhibits semi-Dirac behavior at all energy scales.

\begin{figure}[htb!]
    \centering
    \includegraphics[width=\linewidth]{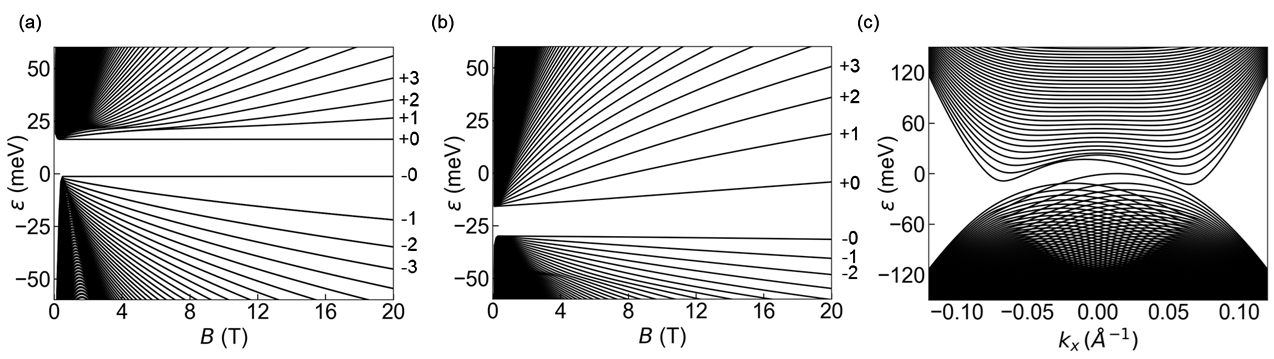}
    \caption{Landau levels of the SrAs$_3$ model when $\bm{B} \parallel \hat{\bm{k}}_x$: (a) $k_x=0$ and (b) $k_x=k_0$. They follow the same scaling behavior as those in Fig. \ref{figs: Landau levels, B//x}, but are gapped by the SOC term.  (c) Landau levels as a function of $k_x$ when $B=10$ T. In (a) and (b), LL indices corresponding to Dirac fermions and semi-Dirac fermions are labeled, respectively.}
    \label{fig:Landau levels, tilt & SOC, B//x}
\end{figure}

{\bf SrAs$_3$ model}: Figure \ref{fig:Landau levels, tilt & SOC, B//x} shows the Landau levels for the SrAs$_3$ model described in Eq. (\ref{eqn: Hamiltonian with tilt and SOC}) when $\bm{B} \parallel \hat{\bm{k}}_x$, including the effects of tilt and SOC. Since the tilt term adds significant complexity to the system, we consider the simplified nodal ring model including the SOC term and analyze the effect of the tilt term for simplicity. Fixing $k_x$, the tilt term can be decomposed into three components: quadratic ($a (k_x^2+k_y^2-k_0^2)$), cross ($a_{xy} k_x k_y$), and constant ($\varepsilon_\text{offset}$) terms, as shown in Eq. (\ref{eqn: simplified + the tilt}). For clarity, we redefine key parameters as $\varepsilon_\text{gas} = b \left( k_x^2 + k_y^2 - k_0^2 \right) $ and $b = \hbar^2 / 2m$:
\begin{equation}\label{eqn: simplified + the tilt}
    H(k_y, k_z) = \left[a (k_x^2+k_y^2-k_0^2)+a_{xy} k_x k_y+\varepsilon_\text{offset}\right]\sigma_0 + b(k_x^2 +k_y^2-k_0^2)\sigma_1 + \hbar v_z k_z \sigma_2 + \Delta_\text{SOC} \sigma_3.
\end{equation}

Figure \ref{fig: tilt, only square} presents the band structures of the simplified nodal ring model in the presence of the quadratic tilt term as a function of $k_y$. The scaling behavior of the energy dispersion remains intact, although the energy dispersion is normalized by the tilt term: linear in Dirac fermions [Figs. \ref{fig: tilt, only square}(a) and \ref{fig: tilt, only square}(b)] and quadratic in semi-Dirac fermions along the $k_y$ direction [Fig. \ref{fig: tilt, only square} (c)].

\begin{figure}[htb!]
    \centering
    \includegraphics[width=\linewidth]{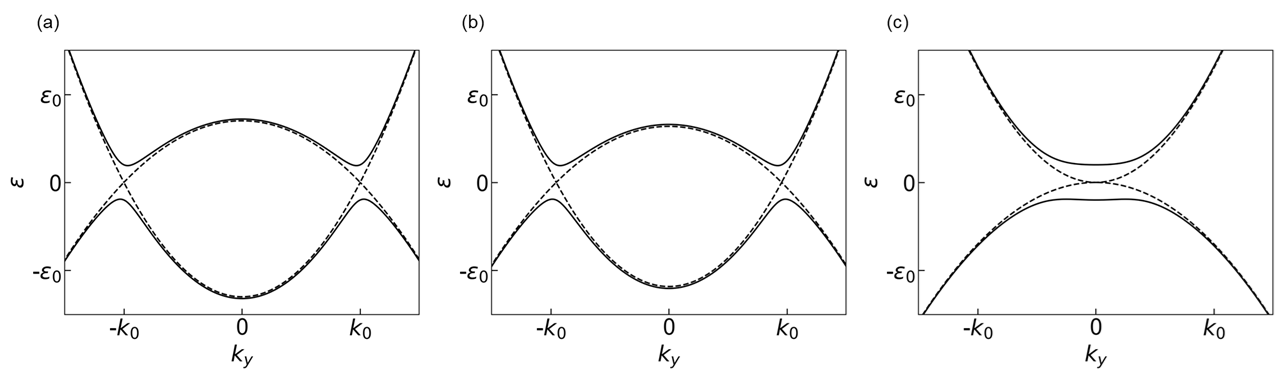}
    \caption{The band structure of the simplified nodal ring model along the $k_y$-direction in the presence of the quadratic tilt term for (a) $k_x=0$, (b) $k_x=0.3 k_0$, and (c) $k_x=k_0$, without the magnetic field. We used $a/b=0.3$, $\Delta_\text{SOC}/\varepsilon_0= 0.2$, and $k_z=0$. Solid and dashed lines represent the band structures with and without the SOC term, respectively. Near the nodal point, (a) and (b) correspond to tilted Dirac fermions, while (c) exhibits semi-Dirac fermions.}
    \label{fig: tilt, only square}
\end{figure}
The energies in the presence of the quadratic tilt term are given by
\begin{equation} \label{eqn: square tilt energy}
\varepsilon = \frac{a}{b} \varepsilon_\text{gas} \pm \sqrt{ \left( \varepsilon_\text{gas} \right)^2 + \left( \hbar v_z \right)^2 + \left( \Delta_{\text{SOC}} \right)^2}.
\end{equation}
After some modifications, similar to Eq. (\ref{eqn: Energy}), we can rewrite Eq. (\ref{eqn: square tilt energy}) as
\begin{equation}
%\varepsilon^2 = \frac{1}{\alpha^4} \left( \varepsilon_\text{gas} + \frac{ab \varepsilon}{b^2-a^2} \right)^2 + \left( \hbar \tilde{v}z k_z \right)^2 + \tilde{\Delta}_\text{SOC}^2,
\varepsilon =\pm\sqrt{ \frac{1}{\alpha^4} \left( \varepsilon_\text{gas} + \frac{ab \varepsilon}{b^2-a^2} \right)^2 + \left( \hbar \tilde{v}z k_z \right)^2 + \tilde{\Delta}_\text{SOC}^2},
\end{equation}
where $\alpha = \sqrt{b^2 / (b^2 - a^2)}$ is a normalization constant. 
The velocity $v_z$ and SOC gap are normalized by the factor $\alpha$, with $\tilde{v}_z = v_z / \alpha$ and $\tilde{\Delta}_\text{SOC} = \Delta_\text{SOC} / \alpha$, but the power-law behavior of the energy dispersion along the $k_y$ and $k_z$ directions remains the same, thus the corresponding scaling behavior of the Landau levels on the magnetic field is also unaffected. A similar normalization in tilted Dirac fermions has been discussed in \cite{Tchoumakov2016, Yu2016, Wyzula2022}.

On the other hand, the cross tilt term shifts the energy with a sign determined by $k_x k_y$.
The energies in the presence of the cross tilt term (neglecting SOC for simplicity) are given by
\begin{equation}
    \begin{aligned}
            \varepsilon &= a_{xy} k_x k_y \pm b \left( k_x^2 + k_y^2 - k_0^2\right)\\
            & = \pm \left[b k_x^2 + b\left(k_y \pm \frac{a_{xy}}{2b}k_x\right)^2 - b k_0^2\right] \mp  \frac{a_{xy}^2}{4b} k_x^2.
    \end{aligned}
\end{equation}
\begin{figure}
    \centering
    \includegraphics[width=\linewidth]{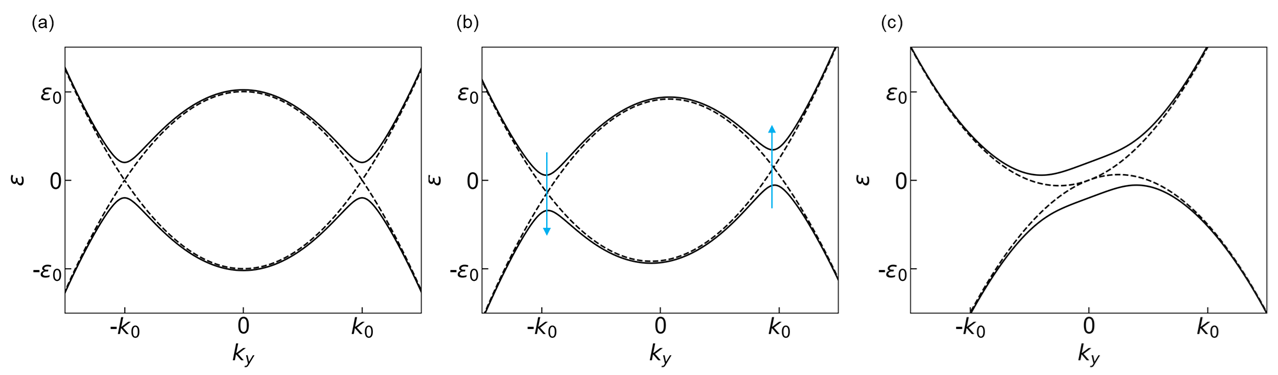}
    \caption{The band structure of the simplified nodal ring model along the $k_y$-direction in the presence of the cross tilt term for (a) $k_x = 0$, (b) $k_x = 0.3k_0$, and (c) $k_x = k_0$, without the magnetic field. We used $a_{xy}/b=0.5$, $\Delta_\text{SOC}/\varepsilon_0 = 0.2$, and $k_z=0$.  Solid and dashed lines represent the band structures with and without the SOC term, respectively. In (b), the directions of the energy shift caused by the cross tilt term for the two Dirac nodes are depicted by blue arrows.}
    \label{fig: tilt, only cross}
\end{figure}
Figure \ref{fig: tilt, only cross} shows the effect of the cross tilt term on the band structures as a function of $k_y$. As $k_y$ moves away from 0, the energies of the two Dirac nodes are shifted oppositely depending on the direction, as illustrated by the blue arrows in Fig. \ref{fig: tilt, only cross}(b), but optical transitions between the corresponding Landau levels are not affected by the cross tilt term. For the semi-Dirac node, however, the cross tilt term modifies the band structure near the node, as illustrated in Fig. \ref{fig: tilt, only cross}(c). Consequently, the corresponding optical transitions between the Landau levels near the semi-Dirac node are also modified, as will be discussed in the next section.

The constant tilt term shifts the energy uniformly, and therefore does not affect optical transitions (except for changing the occupancy of states). Considering the normalization effect of the quadratic tilt term and the energy shift caused by the cross tilt term, the scaling behavior of Landau levels with the magnetic field is preserved.

\clearpage

\section{\uppercase\expandafter{\romannumeral4}. Calculations of the magneto-optical conductivity} \label{Sec: theory, MO}

The magneto-optical conductivity provides insight into the properties of the band structure. Since the gauge field from the light couples to the momentum operator, the linear response to the gauge field can be calculated using the velocity operator $v_i=\partial H / \hbar \partial k_i$.  The magneto-optical conductivity within the Kubo formalism is given by
\begin{equation} \label{eqn: optical conductivity Kubo formula}
\sigma_{ij} (\omega) = -\frac{ie^2}{\hbar} \sum_{n,m} \frac{1}{2 \pi l_B^2}\int{\frac{dk}{2\pi} \frac{f_{n, k}-f_{m, k}}{\varepsilon_{n, k}-\varepsilon_{m, k}} \frac{\left< n, k \left| \hbar v_i \right| m, k \right> \left< m, k \left| \hbar v_j \right| n, k \right>}{\hbar\omega + \varepsilon_{n, k} - \varepsilon_{m, k} + i\eta}}.
\end{equation}
Here, $\varepsilon_{n, k}$ and  $f_{n, k} = 1/\left[  e^{\left( \varepsilon_{n, k} - \mu \right)/k_B T} +1 \right]$ represent the energy and the Fermi-Dirac distribution function, respectively, at a wavevector $k$ and a Landau level index $n$.
$\eta$ is a broadening term, which takes $\eta \rightarrow 0^+$ for the clean limit. For numerical calculations, we used $\eta=4$ meV.  Optical transition peaks arise when $\hbar\omega = \varepsilon_{m, k} - \varepsilon_{n, k}$, corresponding to the energy difference between occupied and unoccupied Landau levels. We extract the scaling behavior of the Landau levels for the magnetic field from the optical transition peaks.

Note that since the reflectivity is derived from the permittivity, which is directly connected to the optical conductivity, \textit{both the magneto-optical conductivity and the magneto-reflectivity exhibit the same Landau-level structures.} \\

\subsection{\uppercase\expandafter{\romannumeral4}-1. Magnetic field parallel to the nodal ring axis ($\bm{B}\parallel\hat{\bm{k}}_z$)}

When a magnetic field aligns parallel to the nodal ring axis, two prominent types of peaks emerge, one stemming from classical fermions and another from disconnected constant energy surfaces. 

{\bf Classical fermions}: Figure \ref{fig: Schematic diagram of optical transition, classical} illustrates the optical transitions occurring between the Landau levels of classical fermions, which approximately form electron and hole gases. As the magnetic field strength increases, the areas enclosed by the constant energy surface expand, consistent with the Onsager quantization rule. The valence Landau levels originate from the electron gas, increasing in energy with the magnetic field, while the conduction levels emerge from the hole gas, decreasing correspondingly. This results in optical transitions between the two Landau levels, which decrease linearly with the magnetic field.

\begin{figure}[htb!]
    \centering
    \includegraphics[width=0.5\linewidth]{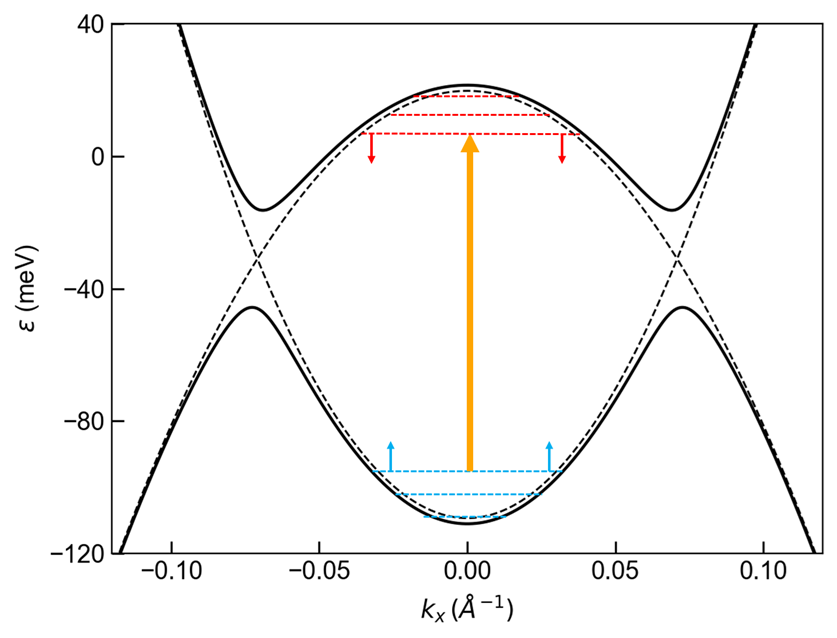}
    \caption{The schematic diagram of the Landau levels and optical transitions alongside the band structure of the SrAs$_3$ model along the direction of $k_x$, with (black solid lines) and without (black dashed lines) the SOC term, respectively, when $k_y=0$ and $k_z=0$. Blue and red dashed lines represent the occupied and unoccupied Landau levels, respectively, with yellow arrows indicating optical transitions. As the magnetic field increases, the Landau levels change in the direction of the blue and red arrows, respectively, increasing the inter-Landau level spacing, thus gradually reducing the optical transition energy linearly with the magnetic field from the band overlap energy $2\varepsilon_0$.}
    \label{fig: Schematic diagram of optical transition, classical}
\end{figure}

When considering the effects of the tilt and SOC terms on the optical transitions between the Landau levels of classical fermions, the tilt term modifies the effective mass. In contrast, the SOC term has a minimal impact, as the corresponding optical transitions occur far above the SOC gap.

Since the SOC term is negligible in the energy range of interest in the experiment, the unoccupied and occupied Landau levels are given by $\varepsilon_+ = \varepsilon_0 - \varepsilon_{+, \text{cl}}$ and $\varepsilon_- = -\varepsilon_0 + \varepsilon_{-, \text{cl}}$, respectively, where $\varepsilon_{\pm, \text{cl}} = \frac{e\hbar}{m_\pm c}(n_\pm + \frac{1}{2})B$ with integer $n_{\pm}$ represents the Landau levels of classical fermions, which are proportional to the magnetic field $B$ and inversely proportional to the effective mass $m_\pm$. 
The optical transition energy is expressed as the difference between the unoccupied and occupied Landau levels, given by
\begin{equation} \label{eqn: cl optical transition}
    \hbar \omega = \varepsilon_+ - \varepsilon_- = 2\varepsilon_0 -(\varepsilon_{+, \text{cl}} + \varepsilon_{-, \text{cl}}).
\end{equation}
Since the Landau level energies of classical fermions are inversely proportional to the effective mass, the tilt term modifies the effective masses of the electron and hole gases, but the corresponding optical transitions still decrease linearly with the magnetic field from the band overlap energy 2$\varepsilon_0=129$ meV.

For fitting, from the optical transition peaks $\hbar\omega$ corresponding to Peak C in Figs. 2(b) and 3(b) of the main text for the experiment and the calculation, respectively, we extracted $\varepsilon_{\pm, \text{cl}}$ in Eq. (\ref{eqn: cl optical transition}) using the energy ratio $\varepsilon_{+,\text{cl}} / \varepsilon_{-, \text{cl}}=m_-(n_++1/2)/m_+(n_-+1/2)$ with the mass ratio $m_+ /m_- = 0.249$ in the SrAs$_3$ model, and obtained the corresponding scaling behavior with $B$ with the LL indices $n_{+}$ and $n_{-}$ involved in the optical transitions, as shown in Fig. 4(b) of the main text. Among $\varepsilon_{\pm, \text{cl}}$, $\varepsilon_{-, \text{cl}}$ is presented in Fig. 4 in the main text.

Regarding the polarization of the electric field, both $\bm{E}\parallel\hat{\bm{k}}_x$ and $\bm{E}\parallel\hat{\bm{k}}_z$ modes exhibit the same scaling behavior for the optical transitions between the classical Landau levels, but the $\bm{E}\parallel\hat{\bm{k}}_x$ mode is less visible compared to the $\bm{E}\parallel\hat{\bm{k}}_z$ mode.
This is because the expectation values of the velocity operator $\left<  n \left| v_i \right| m \right>$ in Eq. (S.13) differ, as $\left<  n \left| v_x \right| m \right> \simeq 0$ and $\left<  n \left| v_z \right| m \right> \neq 0$. This can be simply understood by comparing with the magnitudes of the velocity in the absence of a magnetic field around the origin of the classical fermions: $\left<v_x\right> \simeq 0$, while $\left<v_z\right> \simeq v_z$, originating from $\hbar v_z k_z \sigma_2 s_0$ in Eq. (\ref{eqn: Hamiltonian with tilt and SOC}). \\

{\bf Disconnected constant energy surface}: We analyze the optical transitions involving electron and hole pockets arising from the disconnected constant energy surface, observed just above the SOC gap. Even though both inter-Landau level transitions involving the electron or hole pocket contribute to the magneto-optical conductivity, the dominant contribution originates from transitions involving the electron pocket, as most of the Landau levels of the hole pocket lie above the Fermi energy, which is set at $-15$ meV. This configuration prevents Landau level transitions between unoccupied Landau levels, from the Landau levels of hole pockets to the higher Landau levels.

\begin{figure}[htb!]
    \centering
    \includegraphics[width=0.8\linewidth]{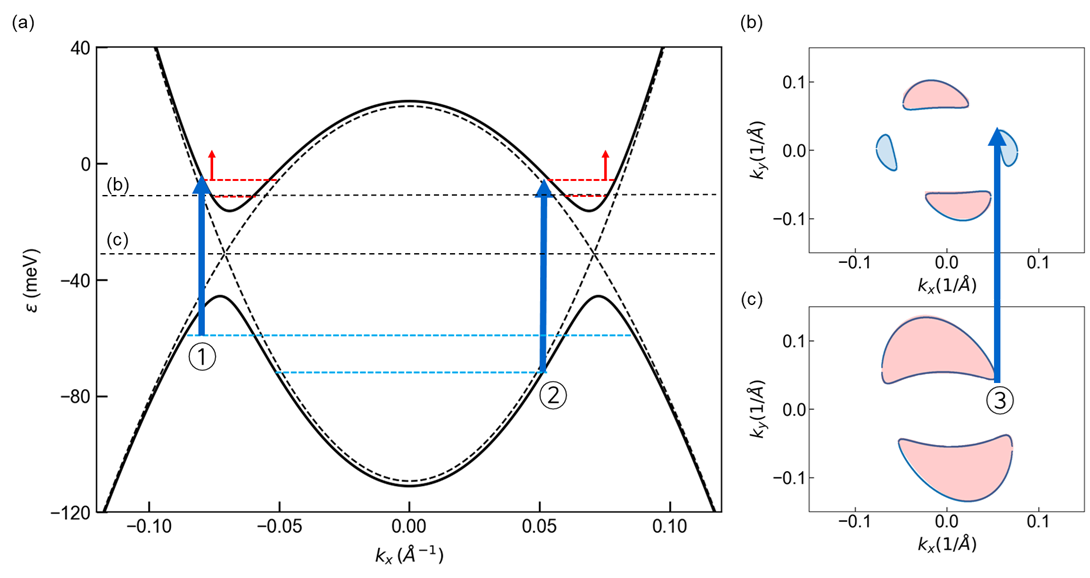}
    \caption{(a) The schematic diagrams of the Landau levels and the band structure of the SrAs$_3$ model, showing optical transitions involving the electron pocket from the hole gas and from the electron gas, with (black solid lines) and without (black dashed lines) the SOC term, respectively. Blue and red dashed lines represent the occupied and unoccupied Landau levels, respectively, with blue arrows indicating optical transitions. 
(b) and (c) show the constant energy surfaces corresponding to (b) the electron pocket at $-10$ meV and (c) the hole pocket at $-30$ meV, represented by dashed lines in (a). The navy arrow represents optical transitions from the hole pockets to the electron pockets.}
    \label{fig: Schematic diagram of optical transition, disconnected FS}
\end{figure}

Figure \ref{fig: Schematic diagram of optical transition, disconnected FS} depicts three types of inter-Landau-level transitions involving the electron pocket, which occur either from the hole gas [\textcircled{1} in Fig. \ref{fig: Schematic diagram of optical transition, disconnected FS}(a)], from the electron gas [\textcircled{2} in Fig. \ref{fig: Schematic diagram of optical transition, disconnected FS}(a)], or from the hole pocket [\textcircled{3} in Figs. \ref{fig: Schematic diagram of optical transition, disconnected FS}(b) and \ref{fig: Schematic diagram of optical transition, disconnected FS}(c)]. Among these, transitions from the electron gas play a dominant role in producing the optical transition peaks associated with the disconnected constant energy surface, observed above the SOC gap. Since optical transitions are restricted to have zero momentum difference, the transitions between the electron pocket and the electron gas are most significant when constant energy surfaces corresponding to two Landau-level energies intersect at the same $k$ point.

The optical transition energies from the disconnected constant energy surface, which are determined by the difference between the Landau levels of the electron pocket and the electron gas, do not represent a specific fermion type and exhibit no scaling behavior associated with it. These optical transition peaks are observed in both $\bm{E} \parallel \hat{\bm{k}}_x$ and $\bm{E} \parallel \hat{\bm{k}}_z$ modes due to the non-zero expectation values of the velocity operator, $\left< n \left| v_i \right| m \right>$.

\subsection{\uppercase\expandafter{\romannumeral4}-2. Magnetic field perpendicular to the nodal ring axis ($\bm{B}\parallel\hat{\bm{k}}_x$)}

When a magnetic field is applied perpendicular to the nodal ring axis, two distinct peaks emerge, one associated with Dirac fermions and the other with semi-Dirac fermions. 

{\bf Dirac fermions}: Figure \ref{fig: Schematic diagram of optical transition, Dirac} depicts the optical transitions between the Landau levels of Dirac fermions. As the magnetic field increases, the energies of the occupied Landau levels decrease, while those of the unoccupied levels increase, both following a 1/2 scaling behavior with the magnetic field.

\begin{figure}[htb!]
    \centering
    \includegraphics[width=0.5\linewidth]{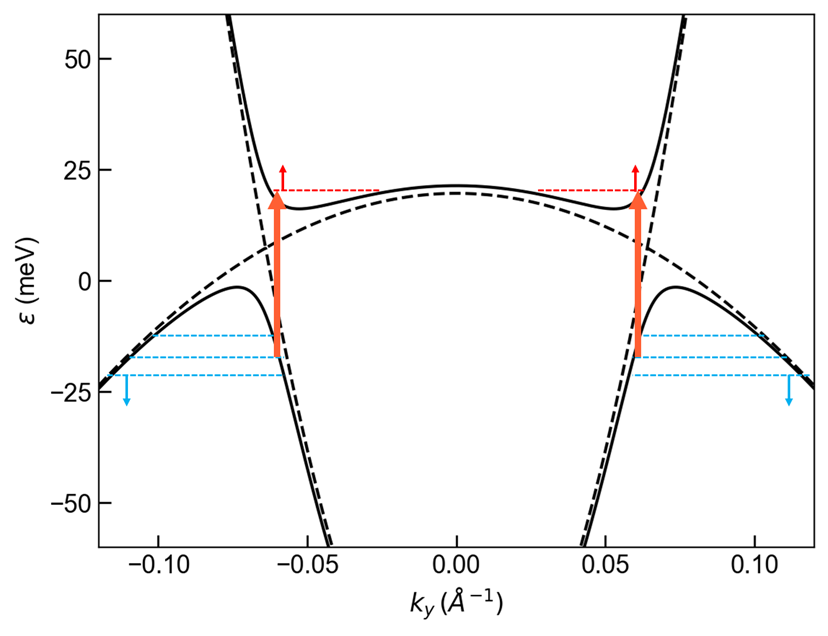}
    \caption{The schematic diagram of the Landau levels and optical transitions alongside the band structure of the SrAs$_3$ model along the direction of $k_y$, with (black solid lines) and without (black dashed lines) the SOC term, respectively,  when $k_x=0$ and $k_z=0$. Blue and red dashed lines represent the occupied and unoccupied Landau levels of the Dirac fermions, respectively, with red arrows indicating optical transitions. As the magnetic field increases, the Landau levels change in the direction of the blue and red arrows.}
    \label{fig: Schematic diagram of optical transition, Dirac}
\end{figure}
The tilt and SOC terms significantly influence the Landau levels by normalizing both their energies and the SOC gap, as previously discussed. The unoccupied and occupied Landau levels measured from the Dirac node are described as $\varepsilon_\pm = \pm  (\varepsilon_{\pm, \text{Dirac}}^2 + \tilde{\Delta}_\text{SOC}^2 )^{1/2}$, where the Landau levels of Dirac fermions follow $\varepsilon_{\pm, \text{Dirac}} = (2 n_\pm e B \hbar \tilde{v}_y \tilde{v}_z /c )^{1/2}$, reflecting the 1/2 scaling behavior with the magnetic field. Here, $\tilde{v}_i = v_i/\alpha$ and $\tilde{\Delta}_\text{SOC}=\Delta_\text{SOC}/\alpha$ represent the normalized velocities and gap, respectively. Note that $\tilde{v}_i$ is the same for the unoccupied and occupied Landau levels. The optical transition energy is given by
\begin{equation} \label{eqn: Dirac, optical transition}
    \hbar \omega =\varepsilon_+ - \varepsilon_- = \sqrt{\varepsilon_{+, \text{Dirac}}^2 + \tilde{\Delta}_\text{SOC}^2}+\sqrt{\varepsilon_{-, \text{Dirac}}^2 + \tilde{\Delta}_\text{SOC}^2}.
\end{equation}
These transition energies begin at the normalized SOC gap $2 \tilde{\Delta}_\text{SOC}$ and increase with the magnetic field with a slope determined by the normalized velocities $\tilde{v}_i$, while $\varepsilon_{\pm, \text{Dirac}}$ still exhibit the same 1/2 scaling behavior with the magnetic field, which is a hallmark of Dirac fermions.

\begin{figure}[htb!]
    \centering
    \includegraphics[width=0.6\linewidth]{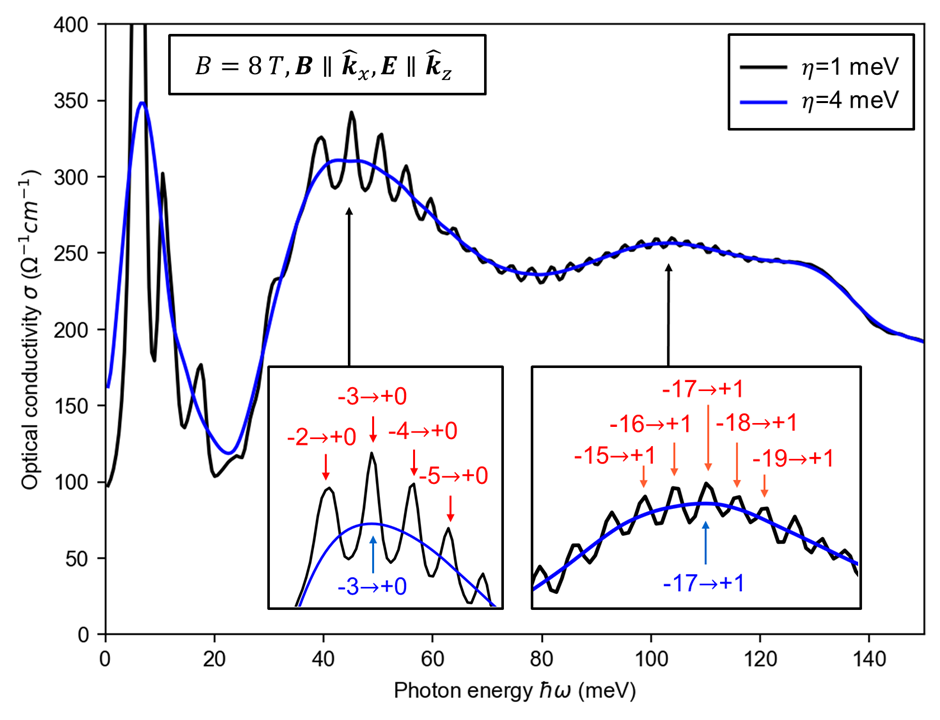}
    \caption{Magneto-optical conductivity at $B=8$ T calculated in the $(\bm{B}, \bm{E}) = (\hat{\bm{k}}_x, \hat{\bm{k}}_z)$ mode when the broadening term $\eta$ is $1$ meV (black) and $4$ meV (blue). The insets show the LL indices involved in the optical transition for Peak A.}
    \label{fig: Peak merge}
\end{figure}

For fitting, from the optical transition peaks $\hbar \omega$ corresponding to Peak A in Figs. 2(a) and 3(a) of the main text for the experiment and the calculation, respectively, we extracted $\varepsilon _{\pm, \text{Dirac}}$ in Eq. (\ref{eqn: Dirac, optical transition}) using the energy ratio $\varepsilon_{+, \text{Dirac}} / \varepsilon_{-, \text{Dirac}}=\sqrt{n_+ / n_-}$ and obtained the corresponding scaling behavior with $B$ with the LL indices $n_{+}$ and $n_{-}$ involved in the optical transitions, as shown in Fig. 4(a) of the main text. Note that for optical transition peaks appearing at high energies, we selected representative $n_+$ and $n_-$ since several LL transitions could be involved in the presence of tilt, as shown in Fig. \ref{fig: Peak merge}. Among $\varepsilon_{\pm, \text{Dirac}}$, $\varepsilon_{-, \text{Dirac}}$ is presented in Fig. 4 in the main text.

Magneto-optical conductivity peaks of Dirac fermions are observed in both polarizations due to nonzero expectation values of the velocity operator $\left<  n \left| v_i \right| m \right>$. Specifically, $\left<  n \left| v_z \right| m \right> \simeq v_z$ arises from the $\hbar v_z k_z \sigma_2 s_0$ term in Eq. (\ref{eqn: Hamiltonian with tilt and SOC}), while $\left<  n \left| v_x \right| m \right> \simeq v_{x, \text{Dirac}}$ which corresponds to the velocity component along the $k_x$-direction of Dirac fermions. \\

{\bf Semi-Dirac fermions}: Figure \ref{fig: Schematic diagram of optical transition, semi-Dirac}(a) illustrates the optical transitions associated with the Landau levels of semi-Dirac fermions. First, assume that the tilt and SOC terms are absent. As the magnetic field increases, the energies of the occupied Landau levels decrease, while those of the unoccupied levels increase, both following a 2/3 scaling behavior with the magnetic field.
Semi-Dirac fermions arise as the elliptical nodal ring becomes tangent to the momentum plane orthogonal to the magnetic field at $k_x = 0.071$ $\rm \AA^{-1}$ in the SrAs$_3$ model, corresponding to the nodal ring radius $k_0$ in the simplified model.
\begin{figure}[htb!]
    \centering
    \includegraphics[width=\linewidth]{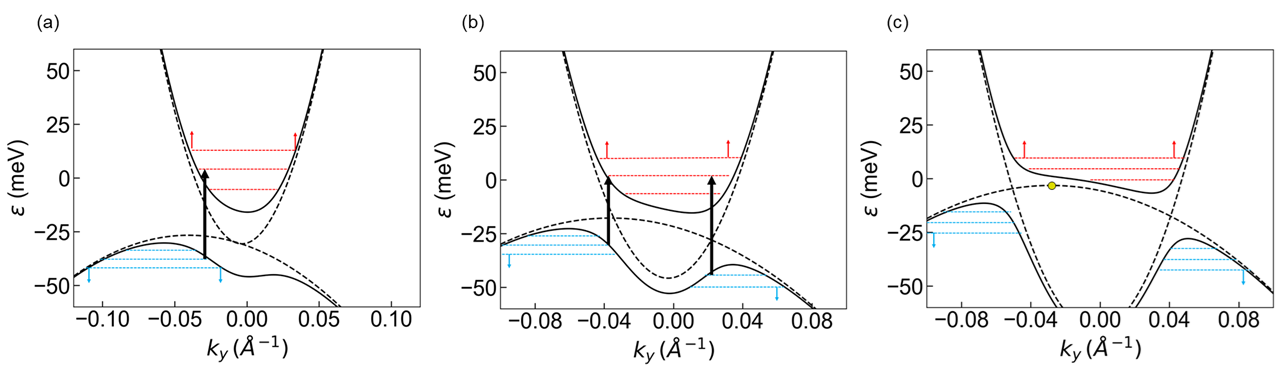}
    \caption{The schematic diagram of the Landau levels and optical transitions alongside the band structure of the SrAs$_3$ model along the direction of $k_y$, with (black solid lines) and without (black dashed lines) the SOC term, respectively, when $k_z=0$ and (a) $k_x=0.071$ $\rm \AA^{-1}$, (b) $k_x=0.064$ $\rm \AA^{-1}$, and (c) $k_x=0.05$ $\rm \AA^{-1}$ . Blue and red dashed lines represent the occupied and unoccupied Landau levels of the semi-Dirac fermions, respectively, with black arrows indicating optical transitions. As the magnetic field increases, the Landau levels change in the direction of the blue and red arrows. The yellow point in (c) illustrates the maximum of the hole gas inside crossing points, exhibiting the Dirac nature.}
    \label{fig: Schematic diagram of optical transition, semi-Dirac}
\end{figure}

When considering the effects of the tilt and SOC terms, the tilt term normalizes the Landau levels and SOC gap, similarly to Dirac fermions. The tilt term also shifts the reference energy levels for semi-Dirac fermions, introducing an offset. The unoccupied and occupied Landau-level energies measured from the maximum of the hole semi-Dirac band in the absence of the SOC are described as $\varepsilon_+ =  (\varepsilon_{+, \text{semi-Dirac}}^2 + \tilde{\Delta}_\text{SOC}^2)^{1/2}-\Delta E$ and $\varepsilon_- =  - (\varepsilon_{-, \text{semi-Dirac}}^2 + \tilde{\Delta}_\text{SOC}^2)^{1/2}$, where the Landau-level energy of the semi-Dirac fermions follows $\varepsilon_{\pm, \text{semi-Dirac}} = [3 \sqrt{\frac{\pi}{2}} \frac{\Gamma(3/4)}{\Gamma(1/4)} \frac{ e\hbar \tilde{v}_z}{\sqrt{\tilde{m}_{y, \pm}} c} \left(n_\pm + 1/2\right) B ]^{2/3}$ \cite{Banerjee2009} exhibiting the 2/3 scaling behavior with the magnetic field. The mass, velocity, and SOC parameters are normalized as $\tilde{m}_{y, \pm} = m_{y, \pm}\alpha^2$, $\tilde{v}_z = v_z/\alpha$, and $\tilde{\Delta}_\text{SOC} = \Delta_\text{SOC}/\alpha$, respectively. Consequently, the optical transition energy becomes
\begin{equation} \label{eqn: semi-Dirac, optical transition}
    \hbar \omega =\varepsilon_+ - \varepsilon_- = \sqrt{\varepsilon_{+, \text{semi-Dirac}}^2 + \tilde{\Delta}_\text{SOC}^2} + \sqrt{\varepsilon_{-, \text{semi-Dirac}}^2 + \tilde{\Delta}_\text{SOC}^2} - \Delta E.
\end{equation}
The optical transition energies increase with the magnetic field with a slope determined by the normalized mass $\tilde{m}_y$ and velocities $\tilde{v}_z$, while $\varepsilon_{\pm, \text{semi-Dirac}}$ preserve the 2/3 scaling inherent to semi-Dirac fermions. 

For fitting, from the optical transition peaks $\hbar \omega$ corresponding to Peak B in Figs. 2(a) and 3(a) of the main text for the experiment and the calculation, respectively, we extracted $\varepsilon_{\pm, \text{semi-Dirac}}$ in Eq. (\ref{eqn: semi-Dirac, optical transition}) using the energy ratio $\varepsilon_{+, \text{semi-Dirac}}/\varepsilon_{-, \text{semi-Dirac}} = \left( \sqrt{\tilde{m}_{y, -}}n_+ / \sqrt{\tilde{m}_{y, +}} n_- \right)^{2/3}$ with the slope ratio $[(\sqrt{\tilde{m}_{y, -}}) / \sqrt{\tilde{m}_{y, +}})]^{2/3}=0.379$ and the energy shift $\Delta E=12.9$ meV in the SrAs$_3$ model, and obtained the corresponding scaling behavior with $B$ with the LL indices $n_{+}$ and $n_{-}$ involved in the optical transitions, as shown in Fig. 4(c) of the main text. Among $\varepsilon_{\pm, \text{semi-Dirac}}$, $\varepsilon_{-, \text{semi-Dirac}}$ is presented in Fig. 4 in the main text.

Here, $\Delta E$ was chosen as the energy shift at $k_x= 0.064 \, \rm \AA^{-1}$, considering the effect of tilt. First, neglect the SOC term. In the presence of tilt, from $k_x= 0.071 \, \rm \AA^{-1}$ to $k_x= 0.064 \, \rm \AA^{-1}$ the single maximum of the hole-like band (which has a heavier mass than that of the electron-like band) appears outside of the crossing points, preserving the semi-Dirac nature  [Figs. S14(a) and S14(b)]. Below $k_x= 0.064 \, \rm \AA^{-1}$, the maximum occurs between the two crossing points, exhibiting the Dirac nature [Fig. S14(c)]. We choose $k_x= 0.064 \, \rm \AA^{-1}$ and the corresponding $\Delta E$ as a representative semi-Dirac point for fitting since this is the point that exhibits the largest joint density of states preserving the semi-Dirac nature. This picture remains valid in the presence of the SOC term.

Regarding the polarization of the electric field, a distinct behavior is observed: the magneto-optical conductivity peaks of semi-Dirac fermions are less visible at low energies in the $\bm{E} \parallel \hat{\bm{k}}_x$ mode than in the $\bm{E} \parallel \hat{\bm{k}}_z$ mode, in contrast to Dirac fermions. This difference arises from the expectation values of the velocity operator $\left< n \left| v_i \right| m \right>$. While $\left< n \left| v_z \right| m \right> \simeq v_z$, contributed by $\hbar v_z k_z \sigma_2 s_0$ in Eq. (\ref{eqn: Hamiltonian with tilt and SOC}), $\left< n \left| v_x \right| m \right> \simeq 0$ near the origin of semi-Dirac fermions.

\subsection{\uppercase\expandafter{\romannumeral4}-3. Details of the numerical calculations}

In the numerical calculations, a finite number of Landau levels are included, with 500 levels used for $3-5$ T and 300 levels for $5-17$ T. These values are chosen to capture the optical transitions within the energy range of interest in the experiment, which is below $150$ meV.

\begin{figure}[htb!]
    \centering
    \includegraphics[width=\linewidth]{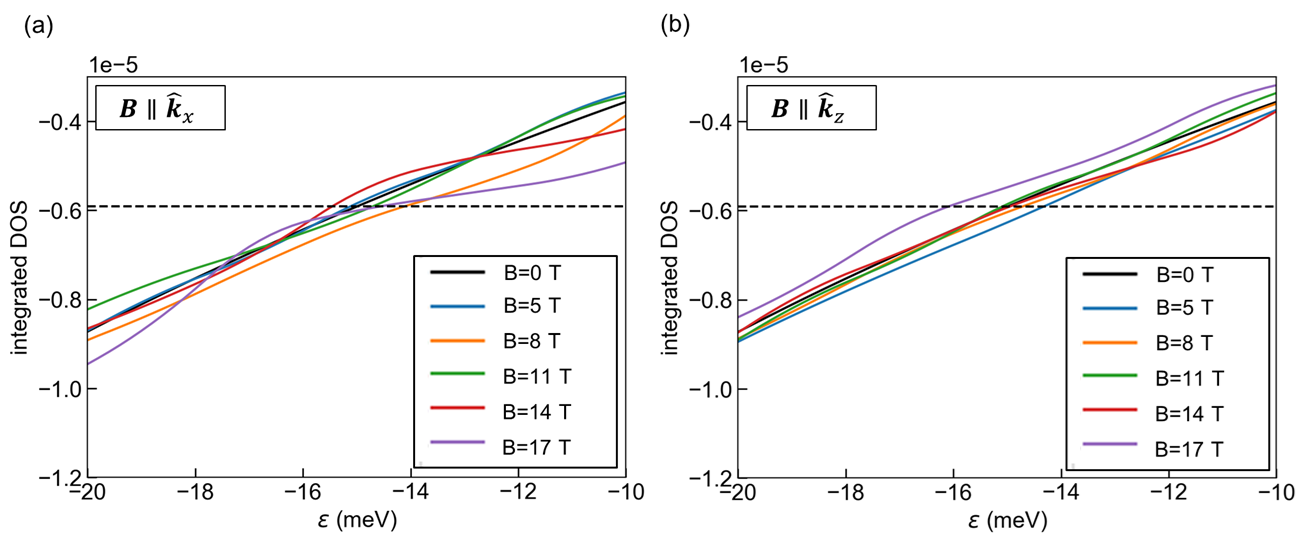}
    \caption{The integrated density of states (DOS) when (a) $\bm{B} \parallel \hat{\bm{k}}_x$ and (b) $\bm{B} \parallel \hat{\bm{k}}_z$. Here, the integrated DOS below $\varepsilon$= 0 is fixed. The dashed lines represent the integrated DOS at $\varepsilon = -15$ meV in the absence of a magnetic field. Even though the magnetic field changes, the Fermi energy shifts only by a few meV.}
    \label{sfig:integrated DOS}
\end{figure}

The Fermi level is fixed at $-15$ meV, independent of the magnetic field. While the application of the magnetic field can slightly shift the Fermi level due to the emergence of Landau levels, calculations of the density of states (DOS) indicate that the shift is minimal, only a few meV as shown in Fig. \ref{sfig:integrated DOS}. Since optical transitions primarily occur above $20$ meV, this small shift in the Fermi level does not affect most of the observed optical transition peaks.

\clearpage

\section{\uppercase\expandafter{\romannumeral5}. Magneto-optical properties for $\bm{E} \parallel \hat{\bm{k}}_x$ mode}
\label{sec:SM-E_kx_mode}

In the main text, we discussed the results for $\bm{E} \parallel \hat{\bm{k}}_z$ modes. Here, we focus on $\bm{E} \parallel \hat{\bm{k}}_x$ modes.  Since the Landau levels are determined by the magnetic field, the scaling behavior with the Landau levels for the magnetic field is independent of light polarization. However, the polarization of the electric field affects the optical transition energy due to different selection rules and the corresponding peak amplitude governed by $\left< n \left| v_i \right| m\right>$ in Eq. (\ref{eqn: optical conductivity Kubo formula}).

\begin{figure}[htb!]
    \centering
    \includegraphics[width=\linewidth]{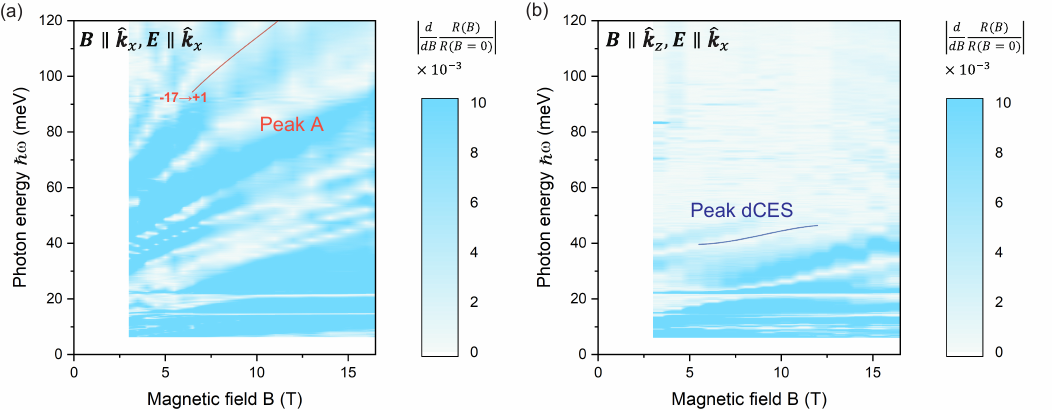}
    \caption{Experimental magneto-optical transition peaks measured in the (a) ($\bm{B}$, $\bm{E}$) $\parallel$ ($\hat{\bm{k}}_x$, $\hat{\bm{k}}_x$) mode and
    (b) ($\bm{B}$, $\bm{E}$) $\parallel$ ($\hat{\bm{k}}_z$, $\hat{\bm{k}}_x$) mode. The derivative $\left| \frac{d}{dB}  \frac{R(B)}{R(B=0)} \right|$ was employed in the 2D color plot to enhance visibility. The shaded thick lines represent the regions exhibiting characteristic Landau level transition peaks corresponding to Peak A-C and Peak dCES distinct from the Landau level transitions.}
    \label{SFig 4: 2d plot E parallel kx experiment}
\end{figure}

Figure \ref{SFig 4: 2d plot E parallel kx experiment} shows the results of magneto-optical reflection measurements. In the ($\bm{B}$, $\bm{E}$) $\parallel$ ($\hat{\bm{k}}_x$, $\hat{\bm{k}}_x$) mode, the LL transitions of Peak A (orange) that arise from the Dirac fermions are distinguishable, while the LL transitions of Peak B (black) that originate from the semi-Dirac fermions are less visible compared to the ($\bm{B}$, $\bm{E}$) $\parallel$ ($\hat{\bm{k}}_x$, $\hat{\bm{k}}_z$) mode as shown in Fig. 2(a) of the main text, especially at low energies. On the other hand, for the ($\bm{B}$, $\bm{E}$) $\parallel$ ($\hat{\bm{k}}_z$, $\hat{\bm{k}}_x$) mode, the LL transitions of Peak C (golden) that arise from the classical fermions are less visible compared to the ($\bm{B}$, $\bm{E}$) $\parallel$ ($\hat{\bm{k}}_x$, $\hat{\bm{k}}_z$) mode as shown in Fig. 2(b) of the main text, while the LL transitions of Peak dCES (blue) that originate from the disconnected constant energy surfaces are still distinguishable. These results are consistent with the theoretical analysis, discussed in Sec. IV.

\begin{figure}[htb!]
    \centering
    \includegraphics[width=\linewidth]{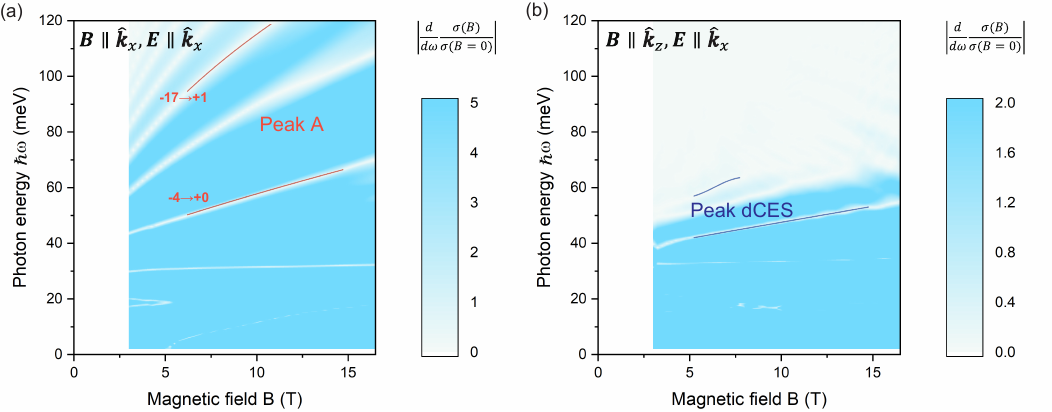}
    \caption{Theoretical magneto-optical transition peaks calculated in the (a) ($\bm{B}$, $\bm{E}$) $\parallel$ ($\hat{\bm{k}}_x$, $\hat{\bm{k}}_x$) mode and (b) ($\bm{B}$, $\bm{E}$) $\parallel$ ($\hat{\bm{k}}_z$, $\hat{\bm{k}}_x$) mode. The derivative $ \left |\frac{d}{d\omega} \frac{\sigma(B)}{\sigma(B=0)} \right|$ was employed in the 2D color plot to enhance visibility. The shaded thick lines represent the regions exhibiting characteristic Landau level transition peaks corresponding to Peak A-C and Peak dCES.}
    \label{SFig 5: 2d plot E parallel kx calculation}
\end{figure}

Figure \ref{SFig 5: 2d plot E parallel kx calculation} illustrates the magneto-optical conductivity derived from theoretical calculations. While experimental reflectivity measurements and theoretical conductivity calculations differ, they both reveal the same peaks linked to optical transitions between Landau levels. Peak A and Peak dCES are clearly distinguishable, whereas Peak B and Peak C are less visible compared with the $\bm{E} \parallel \hat{\bm{k}}_z$ mode, in agreement with the experimental results shown in Fig. \ref{SFig 4: 2d plot E parallel kx experiment}.

\begin{figure}[htb!]
    \centering
     \includegraphics[width=0.6\linewidth]{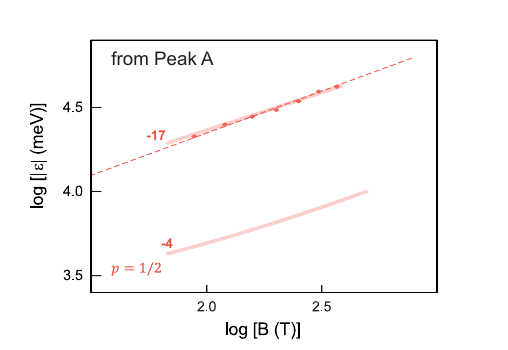}
    \caption{ Logarithmic plots of LL energies as a function of
magnetic field strength for Peak A. Filled circles and solid lines represent experimental and theoretical results, respectively. Dashed lines are used as guides with slopes of 1/2 corresponding to Dirac fermions.}
    \label{Sfig: power laws, E//kx}
\end{figure}

Figure \ref{Sfig: power laws, E//kx} presents the scaling behavior of Landau level energies and the magnetic field, extracted from the optical conductivity peaks, as in Fig. 4 of the main text. Only Peak A is depicted because they are distinctly observed in both experimental and theoretical data. Similar to the $\bm{E} \parallel \hat{\bm{k}}_z$ mode, Peak A exhibits the characteristic scaling behavior of Dirac fermions with a power of 1/2.

\clearpage

\section{\uppercase\expandafter{\romannumeral6}. Comparison with previous magneto-optical experiments on topological semimetals}
\label{sec:SM-Comparison_with_previous_experiments}

\begin{table}[h]
\centering
\caption{\label{table:stable1} The scaling power $\alpha$ of Landau levels for various topological semimetals.}
\setlength{\tabcolsep}{10pt} % Adjust column padding
\renewcommand{\arraystretch}{1.5} % Adjust row spacing
\resizebox{\linewidth}{!}{ % Scale the table to the full width of the page
\begin{tabular}{c|cccc}
\hline
\hline % Top additional line
Material & Topological type & $\alpha$ &
Axis resolved measurements
& Reference\\
\hline
$\rm{TiBiSe}$ & Dirac & $\frac{1}{2}$ & X & \cite{PhysRevB.107.L241101} \\
$\rm{NbP}$ & Weyl & $\frac{1}{2}$ & X & \cite{PhysRevMaterials.6.054204} \\
$\rm{TaAs}$ & Weyl & $\frac{1}{2}$ & X & \cite{doi:10.1126/sciadv.abj1076} \\
$\rm{TaP}$ & Weyl & $\frac{1}{2}$ & X & \cite{PhysRevLett.124.176402} \\
$\rm{BaNiS_2}$ & Nodal line & $\frac{1}{2}$ & X & \cite{PhysRevB.104.L201115} \\
$\rm{NbAs}$ & Nodal line & $\frac{1}{2}$ & X & \cite{PhysRevB.108.L241201} \\
$\rm{ZrSiSe}$ & Nodal line & $\frac{1}{2}$ & X & \cite{Shao2020} \\
$\rm{ZrSiS}$ & Nodal line & $\frac{1}{2}$, $\frac{2}{3}$ & X & \cite{Uykur2019, Shao2024}\\
\arrayrulecolor{red}\hline
\multicolumn{1}{|c}{$\rm{SrAs_3}$} & Nodal ring & $\frac{1}{2}$, 1, $\frac{2}{3}$ & O & \multicolumn{1}{c|}{This work}\\
\hline
\arrayrulecolor{black}\hline % Bottom additional line in black
\end{tabular}
}
\end{table}

In Table \ref{table:stable1}, we summarize previous studies on magneto-optical measurements for various topological semimetals and compare them to our experiments on SrAs$_3$. The Landau levels of topological semimetals exhibit a distinct scaling behavior $\varepsilon \sim B^\alpha$, where the scaling power $\alpha$ depends on their topological nature. We emphasize that only in the current experiment, three types of fermions (classical fermions with $\alpha = \frac{1}{2}$, Dirac fermions with $\alpha = 1$, and semi-Dirac fermions with $\alpha = \frac{2}{3}$) have been observed due to the unique property of SrAs$_3$ possessing a single nodal ring with its curved geometry. We further conducted detailed axis-resolved experiments on SrAs$_3$, allowing a deeper investigation into its unique nodal ring structure and the corresponding topological characteristics.

\clearpage

\bibliographystyle{apsrev4-2}
\bibliography{main}